\documentclass[twocolumn,preprintnumbers,floats,prd,amssymb,floatfix,nofootinbib,balancelastpage,superscriptaddress,amsmath]{revtex4-1}

\pdfoutput=1
\usepackage[utf8]{inputenc}
\usepackage{amssymb}
\usepackage{amsmath}
\usepackage{amsfonts}
\usepackage{graphicx}
\usepackage{color}
\usepackage{xspace}
\usepackage{comment}
\usepackage{hyperref}
\usepackage[normalem]{ulem}
\usepackage[section]{placeins}
\usepackage{afterpage}

\usepackage{float}
\usepackage{slashed}
\usepackage{ulem}

\usepackage{multirow,rotating}
\usepackage[dvipsnames]{xcolor}

\def\gsim{\raise0.3ex\hbox{$\;>$\kern-0.75em\raise-1.1ex\hbox{$\sim\;$}}}
\def\lsim{\raise0.3ex\hbox{$\;<$\kern-0.75em\raise-1.1ex\hbox{$\sim\;$}}}

\newcommand{\ba}[1]{\begin{eqnarray} \label{(#1)}}
\newcommand{\ea}{\end{eqnarray}}

\definecolor{dcolour}{rgb}{.5, .5, .5}

\def\gsim{\raise0.3ex\hbox{$\;>$\kern-0.75em\raise-1.1ex\hbox{$\sim\;$}}}
\def\lsim{\raise0.3ex\hbox{$\;<$\kern-0.75em\raise-1.1ex\hbox{$\sim\;$}}}

\def\m{\rm{\,m}}

\begin{document}

\preprint{APCTP Pre2019-024}

\title{Physics with Far Detectors at Future Lepton Colliders}

\author{Zeren Simon Wang}
\email{zerensimon.wang@apctp.org}
\affiliation{Asia Pacific Center for Theoretical Physics (APCTP) - Headquarters San 31,\\ Hyoja-dong, Nam-gu, Pohang 790-784, Korea}
\affiliation{Physikalisches Institut der Universit\"at Bonn, Bethe Center for Theoretical Physics, \\ Nu{\ss}allee 12, 53115 Bonn, Germany}

\author{Kechen Wang}
\email{kechen.wang@whut.edu.cn}
\affiliation{Department of Physics, School of Science, Wuhan University of Technology,\\ 430070 Wuhan, Hubei, China}

\begin{abstract}
At the Large Hadron Collider (LHC), several far detectors such as FASER and MATHUSLA have been proposed to target the long-lived particles (LLPs) featured with displaced vertices. Naturally one question arises as to the feasibility of installing similar far detectors at future lepton colliders like the CEPC and FCC-ee. Because of the different kinematics of final state particles and the freedom to locate both the experiment hall and the detectors, the future lepton collider with an additional far detector may play a unique role in searching for LLPs. In this study, we consider various locations and designs of far detectors at future $e^- e^+$ colliders  and investigate their potentials for discovering LLPs in the physics scenarios including exotic Higgs decays, heavy neutral leptons, and the lightest neutralinos. Our analyses show that the kinematical distinctions between the lepton and hadron colliders render the optimal positions of far detectors lying at the direction perpendicular to the collider beams at future $e^- e^+$ colliders, in contrast to the LHC where a boost in the forward direction can be exploited. We also find that when searching for LLPs, such new experiments with far detectors at future lepton colliders may extend and complement the sensitivity reaches of the experiments at the future lepton colliders with usual near detectors, and the present and future experiments at the LHC. In particular, we find that, for the theory models considered in this study, a MATHUSLA-sized far detector would give a modest improvement compared to the case with a near detector only at future lepton colliders.
\end{abstract}
\keywords{}


\vskip10mm

\maketitle
\flushbottom
%
%

\section{Introduction}
\label{sec:intro}

In many theories beyond the Standard Model (BSM), new particles are predicted to have a relatively long lifetime; see for instance Refs.~\cite{Essig:2013lka,Alekhin:2015byh,Beacham:2019nyx,Curtin:2018mvb,Alimena:2019zri} for reviews of different models of long-lived particles (LLPs). New particles become long-lived for various reasons including feeble couplings with the standard model (SM) particles, phase space suppression, and heavy mediators. 
Such LLPs, after being produced, travel a macroscopic distance before decaying into other SM and/or new particles. 
Their decays could induce displaced vertices with exotic signatures at colliders. While at the Large Hadron Collider (LHC) most efforts have been focused on searches for promptly decaying new heavy particles, in recent years interests in LLP searches have been growing rapidly. 
Although current LHC experiments with a detector located at the interaction point (which we call ``\textit{near detector}" or abbreviate as ``ND" in this article) have sensitivities to LLPs in some parameter space~\cite{Chatrchyan:2012jna,ATLAS:2012av,Ilten:2015hya,deVries:2015mfw}, a class of new experiments at the LHC with an additional detector located far from the interaction point (which we call ``\textit{far detector}" or abbreviate as ``FD" in this article) have been proposed and shown to have potential sensitivity reaches beyond those of the current LHC experiments in a variety of BSM models.
These proposed experiments include MATHUSLA \cite{Chou:2016lxi,Curtin:2018mvb}, CODEX-b \cite{Gligorov:2017nwh}, FASER \cite{Feng:2017uoz} and AL3X \cite{Gligorov:2018vkc}, which suggest to install an additional detector at a position $\mathcal{O}(5-500)$ m away from different interaction points (IPs) of the LHC.
See Refs.~\cite{Alimena:2019zri,Lee:2018pag} for recent reviews of current and future LLP experiments at the colliders.

While the LHC has recently finished Run 2 and entered Long Shut-down 2 period, discussion of building future colliders has never ceased since decades ago. Several proposals of future colliders including the Circular Electron Positron Collider (CEPC)~\cite{CEPCStudyGroup:2018rmc,CEPCStudyGroup:2018ghi,An:2018dwb} in China, the International Linear Collider (ILC) \cite{Fujii:2017vwa} in Japan, and the Future Circular Collider (FCC) \cite{Abada:2019lih} at CERN have been under investigation for their potential of discovering new physics via performing precision measurements, searching for new heavy particles, etc. 
The ILC would be a linear electron-positron collider, while both the CEPC and FCC would start as circular colliders running with the electron-positron collision mode, which for FCC is called the FCC-ee~\cite{Abada:2019zxq}
\footnote{After the CEPC and FCC-ee come to the end of their operation, the same tunnels are planned to be used for the upgraded proton-proton colliders known as the SppC \cite{CEPC-SPPCStudyGroup:2015esa,Gao:2017ssn} and the FCC-hh \cite{Benedikt:2018csr}, respectively.}.
Compared to hadron colliders such as the LHC, 
lepton colliders running at selected center-of-mass energies ($\sqrt{s}$) can produce a large number of $Z-$, Higgs, and $W-$bosons in clean environments. 
As the intensity frontiers, they would allow not only for precision measurements of SM particles and parameters, but also for discovery of new particles via rare decays of SM particles such as $Z-$ and Higgs bosons.

Inspired by the proposals to construct far detectors at the LHC, in this article we propose similar far detectors could be established at future lepton colliders. 
We consider various locations and designs of far detectors and compare their discovery potentials when searching for LLPs.
As such future lepton colliders are still in the planning period, we focus mainly on the locations (relative to the IP), shapes and volumes of the far detectors, and disregard the concrete details of the availability of space and cost, the technology of detectors and accelerators, etc.

For the physics scenario examples, we consider Higgs decays to a pair of long-lived light scalars at $\sqrt{s} = 240$ GeV, and $Z-$boson decays either to a long-lived heavy neutral lepton (HNL) and an active neutrino, or to a pair of long-lived neutralinos at $\sqrt{s} = 91.2$ GeV.
Some studies associated with the relevant LLPs in these physics scenarios at the CEPC and FCC-ee with near detectors have been performed in Refs.~\cite{Alipour-Fard:2018lsf,Antusch:2016vyf,Wang:2019orr}.

The main purpose of this study is to motivate the construction of far detectors and optimize their basic designs at future electron-positron colliders such as the CEPC and FCC-ee.
The article is organized as follows. 
In Sec.~\ref{sec:fardetectors} we introduce the considered FD designs at $e^- e^+$ colliders.
We then present the basics of the theory models considered in this work in Sec.~\ref{sec:modelbacis}.
The analysis strategies for different physics scenarios, and the results of kinematics and detectors' efficiencies are detailed in  Sec.~\ref{sec:analysisstrategy}.
In Sec.~\ref{sec:numericalresults},  we show the collider sensitivities for three physics scenarios. 
We conclude with a summary of our findings and an outlook in Sec.~\ref{sec:conslusions}.

\section{Far Detector Setups}
\label{sec:fardetectors}

\begin{figure}[]
	\centering
	\includegraphics[width=\columnwidth]{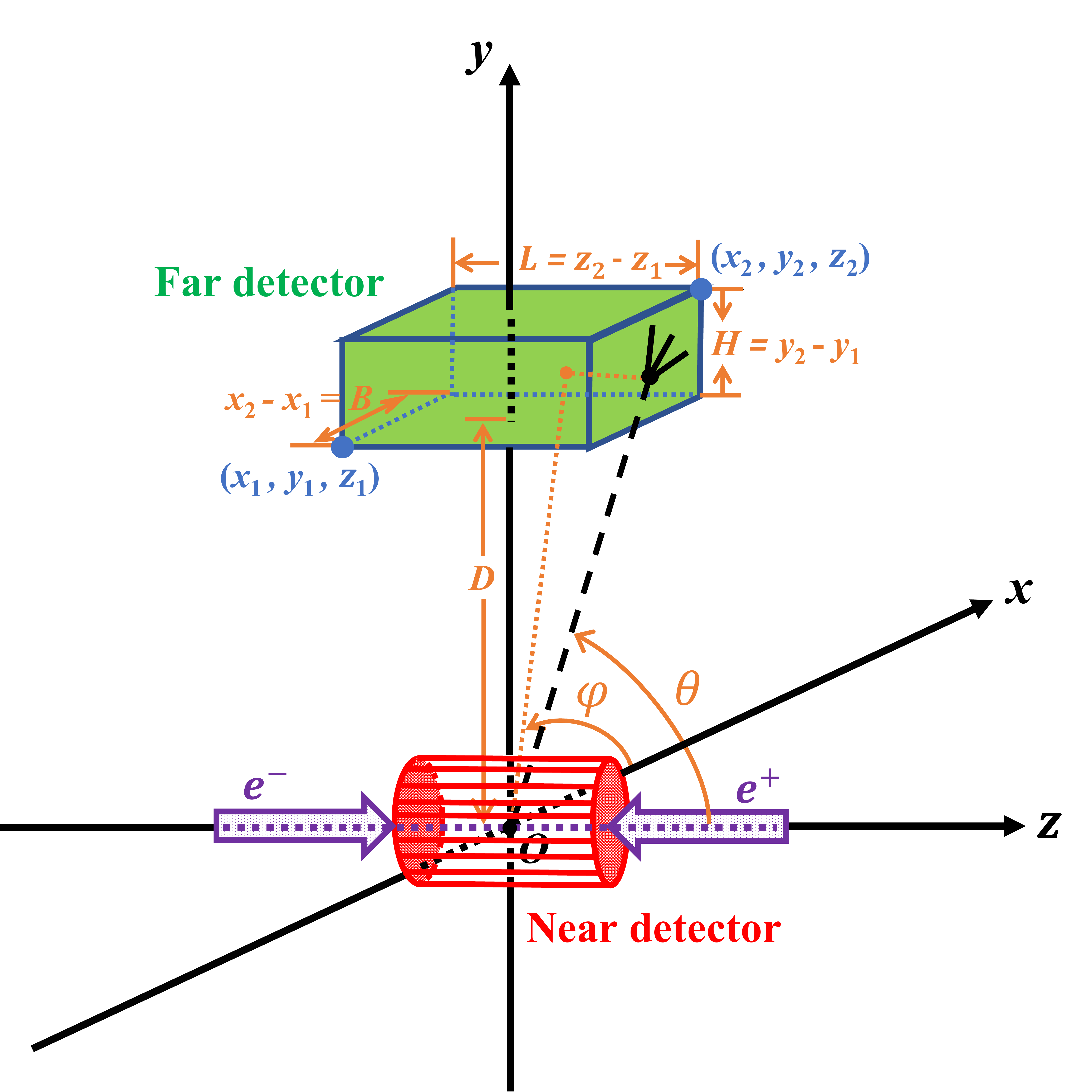}
	\caption{The sketch displays the position of an example far detector. The coordinate system is set up as follows: the origin $O$ is the IP; the injected electron and positron beams travel along the $z$ axis, while the $+z$ direction is defined as the electron beam outgoing direction; the vertical and horizontal axes are set to be $y$ and $x$ axes, respectively;  the $+y$ direction are chosen to be upward. The yellow cylinder enclosing the IP depicts the near detector, while the green cuboid illustrates a far detector located with a distance from the IP.}
	\label{fig:design-sketch}
\end{figure}

In this section, we introduce several possible setups of far detectors (FD1$-$FD8) at future $e^- e^+$ colliders, where we focus on the cuboid shape and consider different installation positions with respect to (w.r.t.) the IP. 
Fig.~\ref{fig:design-sketch} shows a three-dimensional sketch of the colliding beams, the near detector, and a sample far detector.
The coordinate system is set up as follows:
the IP is chosen to be the origin; the $z-$axis is along the incoming electron and positron beams and the ``$+z$" is defined as the electron beam's forward direction; the $x-$axis is the horizontal direction; the $y-$axis is the vertical direction with $+y$ vertically upward.
The polar angle $\theta$ and azimuthal angle $\varphi$ are defined as usual, taking the positive $z-$axis and $x-$axis respectively as their zero values. 
Two coordinates $(x_1,y_1,z_1)$ and $(x_2,y_2,z_2)$ correspond to the two diagonal vertices of the cuboid with the smallest and largest coordinate values, respectively. 
$L = z_2 - z_1$ denotes the length of the detector along $z-$axis; $B = x_2 - x_1$ is its breadth along $x-$axis; and $H = y_2 - y_1$ is its height along $y-$axis. $D$ stands for the radial/transverse distance between the IP and the far detector.

\begin{table*}
	\centering
	\begin{tabular}{c|c|c|c|c|c|c|c|c|c}
	\hline
	\hline
\multirow{2}{*}{\,\,\,\,\,\,\,\,\,\,\,\,\,\,\,}	 & \multirow{2}{*}{$\,\,\,V~[\rm{m^3}]\,\,\,$}  	& \multirow{2}{*}{$B~[\text{m}]$}        & \multirow{2}{*}{$H~[\text{m}]$}       & \multirow{2}{*}{$L~[\text{m}]$}        
& \multirow{2}{*}{$\,\,\,\,(x_1,y_1,z_1)~[\text{m}]\,\,\,\,$}     & \multirow{2}{*}{$\,\,\,\,(x_2,y_2,z_2)~[\text{m}]\,\,\,\,$} & \multirow{2}{*}{$D~[\text{m}]$} 
& \multirow{2}{*}{$\,\,\,\,\,\,\,\,\,\,\,\,\theta~[^\circ]\,\,\,\,\,\,\,\,\,\,\,\,$} & \multirow{2}{*}{$\,\,\,\,\,\,\,\,\,\,\,\,\varphi~[^\circ]\,\,\,\,\,\,\,\,\,\,\,\,$} \\ 
& & & & & & & & & \\
\hline
		\multirow{2}{*}{FD1} & \multirow{2}{*}{$5.0\times10^3$} & \multirow{2}{*}{\,\,\,\,\,10}   & \multirow{2}{*}{10}  & \multirow{2}{*}{\,\,\,\,\,50}   
		                                                 & \multicolumn{1}{c|}{(\,\,\,5, -5, -25)}      &              (15, 5, 25)                             &  \,\,\,\,\,5          &   [11.3, 168.7]              &         [-45.0, 45.0]        \\ \cline{6-10} 
		&               &        &                      &                       &         (10, -5, -25) &                      (20, 5, 25)                             &  \,\,\,10             &   [21.8, 158.2]             &         [-26.6, 26.6]        \\ \hline
		\multirow{2}{*}{FD2} & \multirow{2}{*}{$8.0\times10^5$} & \multirow{2}{*}{\,\,\,200}  & \multirow{2}{*}{20}  & \multirow{2}{*}{\,\,\,200}  
		                                                                            &        (-100, \,\,\,50, \,\,\,50)                    &      (100,  \,\,\,70, 250)      &  \,\,\,50             &   [11.3, \,\,54.5]                &    [26.6, 153.4]           \\ \cline{6-10} 
		&            &           &                      &                       &         (-100, 100, 100)                   &          (100, 120, 300)         & 100            &          [18.4, \,\,50.2]              &     [45.0, 135.0]              \\ \hline
		\multirow{2}{*}{FD3} & \multirow{2}{*}{$8.0\times10^5$}  & \multirow{2}{*}{\,\,\,200}  & \multirow{2}{*}{20}  & \multirow{2}{*}{\,\,\,200}  
		                                                                            &           (-100,  \,\,\,50, -100)                &     (100,  \,\,\,70, 100)       &  \,\,\,50             &  [26.6, 153.4]            &     [26.6, 153.4]            \\ \cline{6-10} 
		&            &           &                      &                       &  (-100, 100, -100)                         &            (100, 120, 100)         & 100            &         [45.0, 135.0]           &     [45.0, 135.0]             \\ \hline
		\multirow{2}{*}{FD4} & \multirow{2}{*}{$8.0\times10^5$}  & \multirow{2}{*}{\,\,\,100}  & \multirow{2}{*}{80}  & \multirow{2}{*}{\,\,\,100} 
		                                                                           &             (-50, \,\,\,50, -50)               &     (50, 130, 50)                  &  \,\,\,50             &  [45.0, 135.0]          &      [45.0, 135.0]            \\ \cline{6-10} 
		&             &          &                      &                       &        (-50, 100, -50)                    &         (50, 180, 50)                 & 100            &          [63.4, 116.6]           &     [63.4, 116.6]            \\ \hline
		\multirow{2}{*}{FD5} & \multirow{2}{*}{$3.2\times10^6$}  & \multirow{2}{*}{\,\,\,200} & \multirow{2}{*}{80} & \multirow{2}{*}{\,\,\,200} &       
		                                                                                       (-100,  \,\,\,50, -100)                    &   (100, 130, 100)           &  \,\,\,50             &   [26.6, 153.4]                &   [26.6, 153.4]  \\ \cline{6-10} 
		&              &         &                      &                       &     (-100, 100, -100)                       &        (100, 180, 100)           & 100                   &   [45.0, 135.0]                  &  [45.0, 135.0]            \\ \hline
		\multirow{2}{*}{FD6} & \multirow{2}{*}{$8.0\times10^7$}  & \multirow{2}{*}{1000} & \multirow{2}{*}{80} & \multirow{2}{*}{1000} 
		                                                                            &     (-500, \,\,\,50, -500)                    &   (500, 130, 500)             &  \,\,\,50             &   [\,\,\,5.7, 174.3]                & [\,\,\,5.7, 174.3]             \\ \cline{6-10} 
		&              &         &                      &                       &     (-500, 100, -500)                       &   (500, 180, 500)             & 100                    &   [11.3, 168.7]                  &     [11.3, 168.7]            \\ \hline
		\multirow{2}{*}{FD7} & \multirow{2}{*}{$8.0\times10^5$}  & \multirow{2}{*}{2000} & \multirow{2}{*}{20}  & \multirow{2}{*}{\,\,\,\,\,20}   
		                                                                            &   (-1000, \,\,\,50, -10)               &    (1000,  \,\,\,70, 10)                  &  \,\,\,50             &   [78.7, 101.3]             &      [\,\,\,2.9, 177.1]           \\ \cline{6-10} 
		&              &         &                      &                       &   (-1000, 100, -10)                      & (1000, 120, 10)                     & 100            &           [84.3, \,\,95.7]           &   [\,\,\,5.7, 174.3]              \\ \hline
		\multirow{2}{*}{FD8} & \multirow{2}{*}{$8.0\times10^5$}  & \multirow{2}{*}{\,\,\,\,\,20}   & \multirow{2}{*}{20}  & \multirow{2}{*}{2000} 
		                                                                           &   (-10,  \,\,\,50, -1000)                  &    (10,  \,\,\,70, 1000)               &  \,\,\,50             &    [\,\,\,2.9, 177.1]        &     [78.7, 101.3]            \\ \cline{6-10} 
		&           &            &                      &                       &  (-10, 100, -1000)                          & (10, 120, 1000)                  & 100            &            [\,\,\,5.7, 174.3]            &   [84.3, \,\,\,95.7]           \\   
\hline
\hline	
	\end{tabular}
	\caption{Summary of proposed far detector designs' geometrical parameters. The detector shapes are assumed to be cuboid with coordinate system sketched in Fig.~\ref{fig:design-sketch}. $V$ stands for the volume of the detector; $B$ its breadth along $x-$axis; $H$ its height along $y-$axis; $L$ its length along $z-$axis; $D$ the radial/transverse distance between the IP and the far detector. $(x_1, y_1, z_1)$ or $(x_2, y_2, z_2)$ are the coordinates of the cuboid's two diagonal vertices with the smallest or largest coordinate values, respectively. $\theta$ is the polar angle covered by the detector, while $\varphi$ is the azimuthal angle.  }
	\label{tab:detectorsparameters}
\end{table*}

In Table~\ref{tab:detectorsparameters}, we list the geometry parameters for FD1--FD8 including the coordinates of the two diagonal vertices and the coverages of the polar angle $\theta$ and azimuthal angle $\varphi$. 
Note that $B$ and $D$ determine the azimuthal angle coverage while $L$ and $D$ determine the polar angle coverage\footnote{For the same $D$ value, FD2 also has a horizontal displacement along the $z-$axis w.r.t. the IP that together with $L$ and $D$ determines the $\theta$-coverage.}, and that $D = x_1$ for FD1 while $D = y_1$ for FD2$-$FD8.

We briefly describe the considered designs of the far detector as follows. 
FD1 is located underground near the IP with its center lying in the same horizontal plane as the IP.
It has the dimensions of $L \times B \times H = 50\m \times 10\m \times 10\m$, and its $y-$ and $z-$dimensions are symmetric relative to the IP ($-5\m < y < +5\m$ and $-25\m < z < +25\m$).
For the horizontal $x-$dimension, there is a 5 or 10 m distance ($D=x_1=5, 10 \m$) from the IP.
This design is considered because there could be some available space near the IP at the future $e^-e^+$ colliders\footnote{FD1 can be placed inside the experiment hall if  the hall is big enough. Otherwise, such kind of detector can be placed in a cavern or shaft near the experiment hall.}.

Depending on the burial depth of the experiment hall, FD2--FD8 lie on the ground with a vertical distance $D = y_1 = $ 50 or 100 m
\footnote{Although currently the experiment hall at the CEPC are assumed to be 100 m below ground, the burial depth of the tunnel and experiment hall can be varied depending on the geological conditions~\cite{CEPCStudyGroup:2018rmc}. Here we take 50 m as one example to demonstrate the case where the depth can be reduced.} 
above the IP, and they are all centrally located in the $x-$dimension. 
FD2 employs the same geometry and a similar relative position to the IP as those of MATHUSLA: 200 m $\times$ 200 m $\times$20 m with a horizontal displacement $z_1 = $ 50 or 100 m w.r.t. the IP. 

In order to understand the effects on the physics sensitivities of the volume, the height, the polar and azimuthal angle coverages, we further propose FD3$-$FD8 which are centrally located in the $z-$dimension.
Among them, FD3 has the same geometry as FD2.
FD4 possesses the same volume as FD3, but its bottom surface area is smaller while its height is larger: 100 m $\times$ 100 m $\times$ 80 m. 
Compared to FD4, FD5 and FD6 have the same height, but their bottom surface areas are much larger: 200 m $\times$ 200 m $\times$ 80 m and 1000 m $\times$ 1000 m $\times$ 80 m, respectively.
Given the unrealistically enormous size, they are studied mainly in order to show the effect of an increased volume.
FD7 and FD8 have the same volume and height as FD3, but employ the shape of a rod: 20 m $\times$ 2000 m $\times$ 20 m and 2000 m $\times$ 20 m $\times$ 20 m, respectively. The difference between FD7 and FD8 is that the rod for FD7 is placed along the $x$-dimension, while it is along the $z$-axis for FD8, so FD7 and FD8 have large $\varphi$ and $\theta$ coverage, respectively. 

By comparing pairs of the far-detector designs, one can find the effects of different geometrical parameters on the sensitivities. For example, for each individual design, we consider two different values of the distance from the IP. 
By comparing FD2 and FD3, one sees the importance of installing a FD centered at zero pseudo-rapidity direction.
Observing FD3 and FD5 helps one understand the effect of changing the height along the $y-$direction, and comparing FD4, FD5, and FD6 shows the effect of varying the length or base surface area.
FD7 and FD8, as two rather extreme designs, can also be compared to FD3 and FD4, in order to see the impact of the ratio between the height and the base surface area.

We also compare the physics sensitivities of the above various far detectors with those of the usual near detectors at the CEPC/FCC.
For the near detectors, the CEPC is equipped with a baseline detector concept \cite{CEPCStudyGroup:2018ghi}. In its inner region, there are a silicon pixel vertex detector, a silicon inner tracker, and a Time Projection Chamber (TPC) which reconstructs the tracks of objects. 
At the FCC-ee, two designs for its near detector have been proposed, namely the ``CLIC-Like Detector'' (CLD) \cite{AlipourTehrani:2254048} and the ``International Detector for Electron-positron Accelerators'' (IDEA)\footnote{The CEPC also takes IDEA as an alternative detector concept. \cite{CEPCStudyGroup:2018ghi}} \cite{Benedikt:2018qee}. As the name says, the CLD design is modified from the CLIC detector after considering the FCC-ee specificities.
Both designs of the FCC-ee's near detector employ a geometrical setup similar to that of the CEPC's baseline detector\footnote{For the rest of this work, we ignore the difference among these near detectors in technology, detector efficiency, etc., and focus on the geometries.}.
Discussion on the near detectors' geometries at the CEPC and FCC-ee can also be found in Ref.~\cite{Wang:2019orr}.
When we present the sensitivity reaches in Sec.~\ref{sec:numericalresults}, the CEPC's baseline detector setup is chosen for the near detector estimate at future lepton colliders. It would lead to almost the identical acceptances, if the FCC-ee's CLD or IDEA detector design is adopted.

\section{Theory Models}\label{sec:modelbacis}
\subsection{Exotic Higgs Decays}\label{subsec:modelHiggs}
The LHC culminated with the discovery of a SM-like Higgs boson in 2012~\cite{Aad:2012tfa,Chatrchyan:2012xdj}. Since then, in order to find (dis)agreement between the observed particle's properties and those predicted by the SM, precision measurements of the Higgs boson have become one of the utmost tasks that the particle physics community are facing.  This is also one of the motivations to build new lepton colliders working as Higgs factories. 

Among all properties, exotic Higgs decays are also important measurements. The exotic Higgs decays into short-lived particles has been studied both at the LHC \cite{Curtin:2013fra,deFlorian:2016spz} and at the future $e^- e^+$ colliders \cite{Liu:2016zki}. Refs.~\cite{Alipour-Fard:2018lsf,Cheung:2019qdr} extended these works to the exotic decay mode of the Higgs to long-lived scalars at future lepton colliders where the authors proposed dedicated search strategies. 

For simplicity, we perform a truth-level study for the physics scenario $h\rightarrow XX$, where $X$ is a new light scalar particle, with both near and far detectors at lepton colliders, and we assume $X$ decays fully visibly (i.e. $\text{Br}(X\rightarrow \text{visible})=100\%$). Our results can be translated to sensitivity reaches in the parameter space of other theoretical models leading to the same or a similar decay topology such as kinetically mixed dark photon \cite{Schabinger:2005ei,Gopalakrishna:2008dv,Curtin:2014cca,Strassler:2008bv}, a lighter Higgs boson in an extended Higgs sector \cite{Curtin:2013fra}, etc.

\subsection{Heavy Neutral Leptons}\label{subsec:modelHNL}
Up to today perhaps the only concrete evidence of BSM physics has been the nonvanishing neutrino mass, confirmed by the neutrino oscillation phenomena. In a class of neutrino seesaw models \cite{Minkowski:1977sc,GellMann:1980vs,Mohapatra:1979ia,Yanagida:1980xy,Schechter:1980gr,Mohapatra:1986aw,Mohapatra:1986bd,Wyler:1982dd,Akhmedov:1995ip,Akhmedov:1995vm, Jana:2018rdf,Das:2019fee}, right-handed sterile neutrinos are added to the SM, explaining the small mass of the active neutrinos via different types of seesaw mechanisms. If such sterile neutrinos exist and are of $\mathcal{O}$(GeV) mass, they may be long-lived with tiny mixings with the active neutrinos. In fact, also known as heavy neutral leptons (HNLs), such particles act as one of the central motivations for LLP studies and for building far detectors at the LHC. 

The HNLs participate in the neutral and charged currents via electroweak interactions described by the following Lagrangian:
\begin{eqnarray}\label{CC-NC}
{\cal L} &=& \frac{g}{\sqrt{2}}\ \sum_{\alpha}
V_{\alpha N}\ \bar \ell_\alpha \gamma^{\mu} P_L N W^-_{L \mu}
+\nonumber \\
&&\frac{g}{2 \cos\theta_W}\ \sum_{\alpha, i}V^{L}_{\alpha i} V_{\alpha N}^*
\overline{N} \gamma^{\mu} P_L \nu_{i} Z_{\mu},
\end{eqnarray}
where $i=1,2,3$, and $\ell_\alpha$ ($\alpha=e,\,\mu, \tau$) are the charged leptons of the SM. 
For simplicity, we study the case where only one HNL ($N$) mixes with one single generation of active neutrinos $\nu_\alpha$ for $\alpha=e / \mu$, and treat the mass of HNL $m_N$ and the mixing parameters $|V_{\alpha N}|^2$ between the $N$ and active neutrinos $\nu_{\alpha}$ as free parameters.

At an $e^- e^+$ collider, a HNL could be produced via an $s-$channel $Z-$boson or via a $t-$channel $W-$boson exchange\footnote{In the latter case, the HNL can only be mixed with the electron neutrino.}. A detector-level study of displaced vertices of HNLs at future lepton colliders has been performed in Ref.~\cite{Antusch:2016vyf}, where the authors points out that the CEPC running with $\sqrt{s}=91.2$ and 250 GeV could have the sensitive reaches in $m_N$ up to almost 80 GeV. In this study, we focus on the HNLs produced from $Z-$decays with a general $e^- e^+$ collider running at the $Z-$pole (with $\sqrt{s}=91.2$ GeV), and provide sensitivity predictions for both the far detectors FD1$-$FD8 and near detectors at the CEPC and FCC-ee. 

The decay width of a $Z-$boson into an active neutrino and a HNL is calculated with:
\begin{eqnarray}
\Gamma(Z\rightarrow  N && \nu_{\alpha}) = 2\cdot \Gamma(Z\rightarrow \nu_\alpha \bar{\nu}_\alpha) \cdot |V_{\alpha N}|^2   \nonumber\\
&&   \cdot \big( 1 - (m_N/m_Z)^2 \big)^2 \big( 1+\frac{1}{2}(m_N / m_Z)^2 \big).\label{eqn:Z2nuNdecaywidth}
\end{eqnarray}
We assume the neutrinos are of Majorana nature (hence the factor 2 in Eq.~\eqref{eqn:Z2nuNdecaywidth}) and calculate their decay widths with the formulas given in Ref.~\cite{Atre:2009rg}. In addition, we take into account only the visible branching ratios of the HNLs (i.e. we include all the decay channels except the fully invisible tri-neutrino one).

\subsection{R-parity-violating Supersymmetry and the Light Neutrailnos}\label{subsec:modelRPV}

Light neutralinos of mass $\mathcal{O}$(GeV) are still allowed by both observational and laboratory measurements \cite{Choudhury:1999tn,Hooper:2002nq,Bottino:2002ry,Dreiner:2009ic,Vasquez:2010ru,Calibbi:2013poa,Dreiner:2003wh,Dreiner:2013tja}, as long as they can decay with a lifetime much shorter than the age of the Universe. One possibility to realize such decays is to have RPV-SUSY (see Refs.~\cite{Barbier:2004ez,Dreiner:1998wm,Mohapatra:2015fua} for reviews). 

The most general form of RPV superpotential can be written as follows:
\begin{eqnarray}
W_{\text{RPV}}&=&\mu_i H_u \cdot L_i + \frac{1}{2}\lambda_{ijk}L_i \cdot L_j \bar{E}_k \nonumber\\
&& + \lambda'_{ijk} L_i \cdot Q_j \bar{D}_k + \frac{1}{2}\lambda''_{ijk}\bar{U}_i \bar{D}_j \bar{D}_k,
\end{eqnarray}
where the first three terms lead to lepton number violation and the last set of operators violate baryon number. If the neutralinos are light and the RPV couplings are small but nonvanishing, the lightest neutralinos may become long-lived and result in exotic signatures such as displaced vertices at colliders. 

In this study, for illustration purpose we consider only one nonvanishing operator, $\lambda'_{112}L_1 Q_1 \bar{D}_2$, and study the potential sensitivities of various far-detector designs at a general $e^- e^+$ collider. We estimate the discovery reaches in the parameter space of $\lambda'_{112}/m^2_{\tilde{f}}$ vs. $m_{\tilde{\chi}_1^0}$, where $m^2_{\tilde{f}}$ is the sfermion mass squared, and all sfermions' masses are assumed to be degenerate. 

As described in Ref.~\cite{Dercks:2018wum}, for $m_{\tilde{\chi}_1^0}\lesssim 3.5$ GeV the light neutralinos would undergo mainly two-body decays into a lepton/neutrino and a charged/neutral meson, while for larger mass values, three-body decays would be dominant. We calculate the two-body decay widths using the analytic expressions given in Ref.~\cite{deVries:2015mfw}, and the three-body decay widths by the program SPheno-4.0.3 \cite{Porod:2003um,Porod:2011nf}.
We also assume all the lightest neutralinos ${\tilde{\chi}_1^0}$ can be identified regardless of their decay modes in this study (i.e.  $\text{Br}(X\rightarrow \text{visible})=100\%$ in Eq.~(\ref{eqn:LLPdecay})).
Note that ${\tilde{\chi}_1^0}$  decays to charged lepton final state with a branching ratio $\sim 0.5$, which leads to a very small reduction in the sensitivity reaches if only the charged lepton final state is considered.

We consider the light neutralinos produced in pair from on-shell $Z-$boson decays. Light $\mathcal{O}$(GeV) neutralinos consist dominantly of a bino, with only a small component of Higgsinos. While it is only the latter coupled to a $Z-$boson, the large number of $Z-$bosons produced at future $Z-$factories such as the CEPC could still offset the necessarily small branching ratio of $Z\rightarrow \tilde{\chi}_1^0\tilde{\chi}_1^0$. 

As discussed in detail in Ref.~\cite{Helo:2018qej}, the ATLAS experiment at the LHC set a lower limit on the Higgsino mass $\mu \geq 130$ GeV \cite{Aaboud:2017leg} which can be translated via supersymmetry theoretical calculation into an upper bound on Br$(Z\rightarrow  \tilde{\chi}_1^0\tilde{\chi}_1^0)\lesssim 0.06\%$ for $m_{\tilde{\chi}_1^0}\ll m_Z/2$.
This limit almost saturates the experimental upper bound on Br$(Z \rightarrow \tilde{\chi}_1^0\tilde{\chi}_1^0)\sim 0.1 \%$ derived from the LEP measurement of the invisible width of the $Z-$boson \cite{Patrignani:2016xqp}. In this study, we choose the benchmark value of Br($Z\rightarrow\tilde{\chi}_1^0\tilde{\chi}_1^0)=10^{-3}$, and treat $m_{\tilde{\chi}_1^0}$ and $\lambda'_{112}/m^2_{\tilde{f}}$ as free parameters.\footnote{The branching ratio numbers are for $m_{\tilde{\chi}_1^0}\ll m_Z/2$. For larger $m_{\tilde{\chi}_1^0}$, phase space effects are taken into account in our MC simulation.} 

\section{Analysis strategy}
\label{sec:analysisstrategy}

\subsection{Physics Scenarios}

In this study, we consider the CEPC and FCC-ee as the benchmark lepton colliders. 
At $\sqrt{s}=240$ GeV, the Higgs bosons are produced via three processes: $e^- e^+ \rightarrow Z h$ via a virtual $Z-$boson (dominantly), and two vector boson fusion (VBF), i.e. $WW-$ and $ZZ-$fusion, processes. 
Since the CEPC is planned to be operated as such a Higgs factory for 7 years with two IPs, it is expected to produce a total number of $1.14\times 10^6$ Higgs bosons which correspond to an integrated luminosity of 5.6 ab$^{-1}$~\cite{CEPCStudyGroup:2018ghi}.
For the FCC-ee, fewer years are planned for the Higgs factory mode. 
However, its larger instantaneous luminosity renders roughly a million Higgs bosons to be produced as well. 
Therefore, in this study, we specify the total number of the SM Higgs bosons produced at either the CEPC or FCC-ee as $N_h=1.14\times 10^{6}$ with $\mathcal{L}_h=5.6$ ab$^{-1}$. 

As a $Z-$factory with $\sqrt{s}=91.2$ GeV, the CEPC would collect almost one trillion $Z-$bosons which are dominantly produced via the $e^- e^+ \rightarrow Z$ process in 2 years with two IPs, i.e. $N_Z^{\mathrm{CEPC}}=7.0 \times 10^{11}$ corresponding to a total integrated luminosity of $\mathcal{L}_{Z}^{\mathrm{CEPC}}=16$ ab$^{-1}$~\cite{CEPCStudyGroup:2018ghi}.
The FCC-ee will also run at the $Z-$pole, producing $N_Z^{\mathrm{FCC-ee}}=5.0 \times 10^{12}$ $Z-$bosons ($\mathcal{L}_Z^{\text{FCC-ee}}=150$ ab$^{-1}$) in 4 years with two IPs~\cite{Abada:2019zxq}.
It is worth noting that the values of $N_h$ and $N_Z^{\text{CEPC}/\text{FCC-ee}}$ we use in this work are associated with two IPs at either the CEPC or FCC-ee, and, correspondingly, two identical FDs are required (with one above each IP) for the results shown in the present study.
In case there is only one FD installed, the total available integrated luminosity will be reduced by a factor of 1/2. Sensitivity reaches on the model parameters would be reduced slightly, not affecting the results qualitatively.

\begin{table}[]
\begin{center}
\begin{tabular}{|c|c|c|c|c|}
\hline
\multicolumn{2}{|c|}{scenario}  & $h\rightarrow XX$ & $\,\,Z\rightarrow N \nu\,\,$  & $\,\,Z\rightarrow \tilde{\chi}_1^0\tilde{\chi}_1^0\,\,$ \\
\hline
\multicolumn{2}{|c|}{LLP}  & $X$ & $N$  & $\tilde{\chi}_1^0$ \\
\hline
\multicolumn{2}{|c|}{production} & $Zh$ (main)  & \multicolumn{2}{c|}{\multirow{2}{*}{$Z$}} \\
\multicolumn{2}{|c|}{$e^-e^+\to$} & $\nu\bar{\nu}h, e^-e^+h$ (VBF) & \multicolumn{2}{c|}{} \\

\hline
\multicolumn{2}{|c|}{$\sqrt{s}$ [GeV]}  & 240  & \multicolumn{2}{c|}{91.2} \\
\hline
\multirow{2}{*}{$N_h$} & CEPC    & \multirow{2}{*}{$1.14\times 10^{6}$~\cite{CEPCStudyGroup:2018ghi}} & \multicolumn{2}{c|}{\multirow{2}{*}{-}} \\
                                        & FCC-ee & &  \multicolumn{2}{c|}{}  \\
\hline
\multirow{2}{*}{$N_Z$} & CEPC    & \multirow{2}{*}{-}  & \multicolumn{2}{c|}{$7.0 \times 10^{11}$~\cite{CEPCStudyGroup:2018ghi}} \\
                                        & FCC-ee & & \multicolumn{2}{c|}{$5.0 \times 10^{12}$~\cite{Abada:2019zxq}} \\ 
\hline
\end{tabular}
\caption{Summary of the considered physics scenarios for the LLPs and the production modes of their parent particles at the CEPC and FCC-ee with center-of-mass energy $\sqrt{s} = $ 240 and 91.2 GeV, respectively. The numbers of Higgs and $Z-$bosons ($N_h$ and $N_Z$) at different colliders used in this study are listed in the last four rows.}
\label{tab:physicscases}
\end{center}
\end{table}

We choose to investigate the three physics scenarios discussed in the previous section, where the LLPs are produced from Higgs or $Z-$bosons decays. 
We firstly consider the Higgs bosons decaying into a pair of light scalars $X$: $h\rightarrow XX$, varying the proper decay length $(c\tau)$ of $X$ and Br($h\rightarrow XX$) to find the sensitive parameter spaces for various far detector designs. 
For $Z-$boson decays, we take the HNLs, $N$, and the lightest neutralinos, $\tilde{\chi}_1^0$, as the benchmark scenarios, and estimate detectors'  sensitivities in their respective theoretical parameter space. 
The relevant information is summarized in Table~\ref{tab:physicscases}.
We note that LLPs with mass around GeV scale could also be produced from rare meson decays. Although it is not included in this study, the sensitivity reaches in the low mass region could be enhanced if such production is added.

\subsection{Signal Simulations}

We calculate the total number of LLPs produced, $N_{\text{LLP}}^{\text{prod}}$, with the following expression:
\begin{eqnarray}
N_{\text{LLP}}^{\text{prod}} = \sum_M N_M\cdot n_{\text{LLP}}\cdot  \text{Br}(M\rightarrow n_{\text{LLP}} \,\text{LLP}+Y),
\end{eqnarray}
where the summation $\sum$ is performed over different types of the mother particle $M$ of the LLPs (i.e. Higgs or $Z-$boson); $N_M$ denotes the total number of $M$; $n_{\text{LLP}}=1,2,\ldots$ is the number of the LLP(s) produced from each $M$ decay; and $Y$ represents other particles associated with LLP(s) production if any. In this study, Br($M\rightarrow n_{\text{LLP}} \,\text{LLP(s)}+Y$) is treated as an independent parameter for the physics scenarios of $h\rightarrow XX$ and $Z\rightarrow \tilde{\chi}_1^0\tilde{\chi}_1^0$, and is calculated analytically for $Z\rightarrow N \nu$. 

We proceed to determine the average decay probability of the LLPs inside the detector fiducial volume, $\left\langle P[\text{LLP in f.v.}]\right\rangle$, with the following formula:
\begin{eqnarray}
\langle P[\text{LLP}\text{ in f.v.}]\rangle=\frac{1}{N^{\text{MC}}_{\text{LLP}}}\sum_{i=1}^{N^{\text{MC}}_
{\text{LLP}}}P[(\text{LLP})_i\text{ in f.v.}]\,,
\label{eq:decay_prob}
\end{eqnarray}
where $N^{\text{MC}}_{\text{LLP}}$ is the total number of LLPs generated with the MC simulation tool Pythia 8.205 \cite{Sjostrand:2006za,Sjostrand:2014zea}, and $P[(\text{LLP})_i\text{ in f.v.}]$ denotes the decay probability of an individual LLP inside the decay chamber. In order to calculate the latter, we extract the kinematics of the physics processes from Pythia and make use of exponential decay law. 

For the Higgs-boson generation, we apply the ``HiggsProcess'' module where we turn on the ``HiggsSM:all'' switch taking into account all the three Higgs production ($HZ$, $WW-$ and $ZZ-$fusion) processes mentioned above. We set the Higgs bosons to decay solely into a pair of new scalars in order to obtain the maximal number of statistics.

For the SM $Z-$boson simulation, since the $Z-$boson is hard-coded to have the SM properties in Pythia, it cannot easily be set to decay into new particles with a branching ratio. We thus use the ``New-Gauge-Boson'' module to generate $Z'-$bosons. 
We tune the mass and couplings of $Z'-$boson to be the same as those of the SM $Z-$bosons and demand that these $Z'-$bosons decay only into either a new fermion plus a $\nu$, or a pair of new fermions to obtain the kinematics of the $Z\rightarrow N \nu$ or $Z \rightarrow \tilde{\chi}_1^0 \tilde{\chi}_1^0$ processes, respectively. 

The calculation of the individual decay probability $P[(\text{LLP})_i\text{ in f.v.}]$ depends on the detector's geometries and its position relative to the IP.
With the average decay probability expressed, we calculate the total number of LLPs decaying in the fiducial volume as:
\begin{eqnarray}
N_{\text{LLP}}^{\text{obs}}=N_{\text{LLP}}^{\text{prod}} \cdot \left\langle P[\text{LLP in f.v.}]\right\rangle \cdot \text{Br(LLP}\rightarrow\text{visible}),\nonumber\\
\label{eqn:LLPdecay}
\end{eqnarray}
where Br(LLP $\rightarrow$ visible) denotes the decay branching ratio of the LLP into visible final state.
The latter depends on the model parameters and the properties of the corresponding LLP, which have been discussed in Sec.~\ref{sec:modelbacis}.
This factor is included to ensure that the secondary vertex could be reconstructed.

Based on the kinematic information of each LLP provided by Pythia, we can derive the kinematic variables as follows:
\begin{eqnarray}
\beta_i^z &=& p_i^z/E_i,\\
\gamma_i &=& E_i/m,\\
\lambda_i^z &=& \beta_i^z \, \gamma_i \, c \, \tau,
\end{eqnarray}
where $p_i^z$ is the $z-$momentum of $(\text{LLP})_i$, $E_i$ ($m$) is its energy (mass), and $c\tau$ is its proper decay length. 

The following formulas are then used to calculate the decay probability of an individual LLP inside each far detector design:
\begin{eqnarray}
P[(\text{LLP})_i{\text{ in FD2}}]&=&\frac{2\arctan{(B/(2D))}}{2\pi} \frac{1-e^{-S'_2/\lambda_i^z}}{e^{S_2/\lambda_i^z}}\,,\label{eqn:individualdecprob2}\\
S_2&\equiv&\text{min}\bigg(\text{max}\big(D,\frac{D}{\tan{\theta_i}}\big),D+L \bigg)\,\nonumber,\\
S'_2&\equiv&\text{min}\bigg(\text{max}\big(D,\frac{D+H}{\tan{\theta_i}}\big),D+L\bigg)-S_2\,,\nonumber\\
P[(\text{LLP})_i{\text{ in FDj}}] &=& \frac{2\arctan{(B/(2D))}}{2\pi} \frac{1-e^{-S'_j/\lambda^z_i}}{ e^{S_j/\lambda_i^z}}\,,\label{eqn:individualdecprob134567}\\
S_j &\equiv&  \text{min}\left(\frac{L}{2},\bigg|\frac{D}{\tan \theta_i }\bigg|\right),\nonumber\\
S'_j &\equiv& \text{min}\left(\frac{L}{2}, \bigg|\frac{D+H}{\tan{\theta_i}}\bigg|\right)-S_j. \nonumber
\end{eqnarray}
Here Eq.~\eqref{eqn:individualdecprob2} is for FD2, while Eq.~\eqref{eqn:individualdecprob134567} is for the other far detectors. 
$\theta_i$ denotes the polar angle of (LLP)$_i$.
The formula for FD2 is different because only FD2 has a horizontal displacement w.r.t. the IP. The prefactors $(2\arctan{(B/(2D))})/2\pi$ in both of Eq.~\eqref{eqn:individualdecprob2} and Eq.~\eqref{eqn:individualdecprob134567} account for the coverage of the azimuthal angle $\varphi$ assuming flat differential distribution of the LLPs in $\varphi$, and the remaining exponential factors take into account the polar angle coverage and `fiducial length' of the detectors along $z-$axis. These factors, combined with the boosted decay length in the $z-$direction, allow for obtaining the individual decay probabilities.

\subsection{Kinematical Distributions}

\begin{figure}[]
	\centering
	\includegraphics[width=\columnwidth]{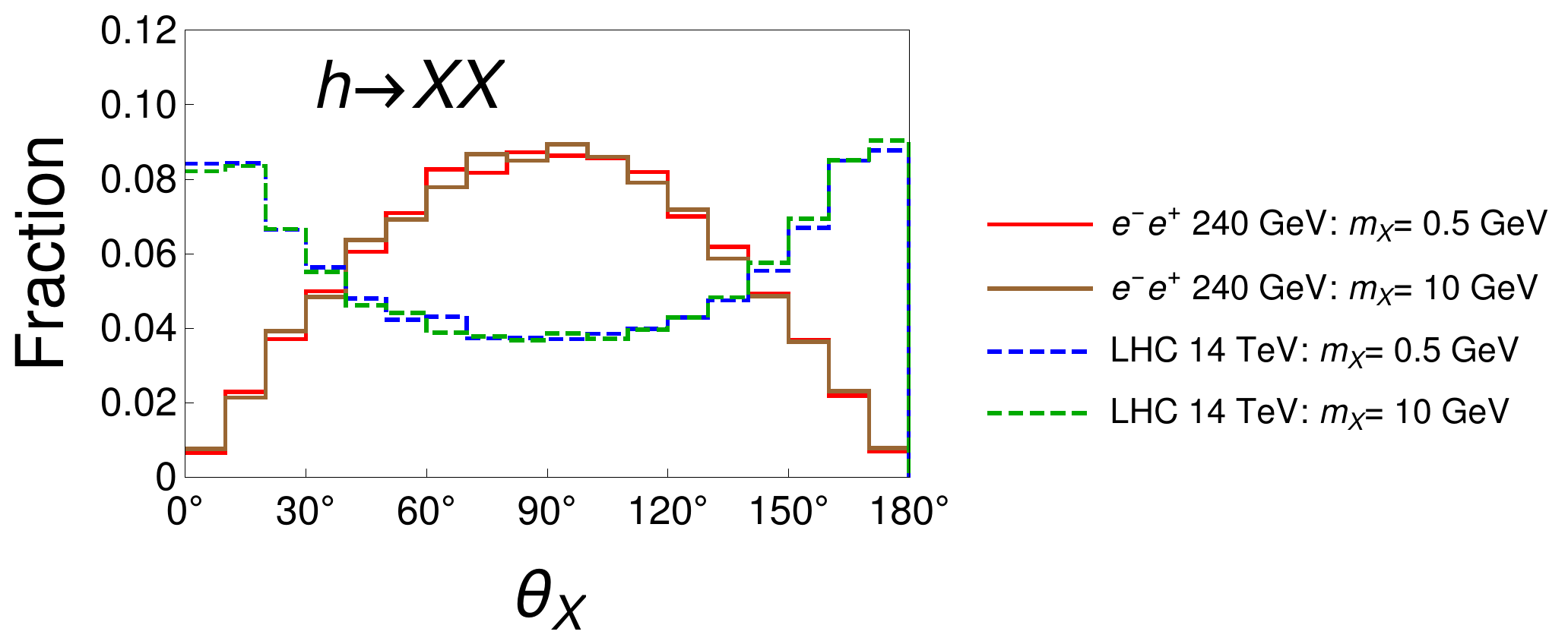}
	\par \vspace{0.3cm}
	\includegraphics[width=\columnwidth]{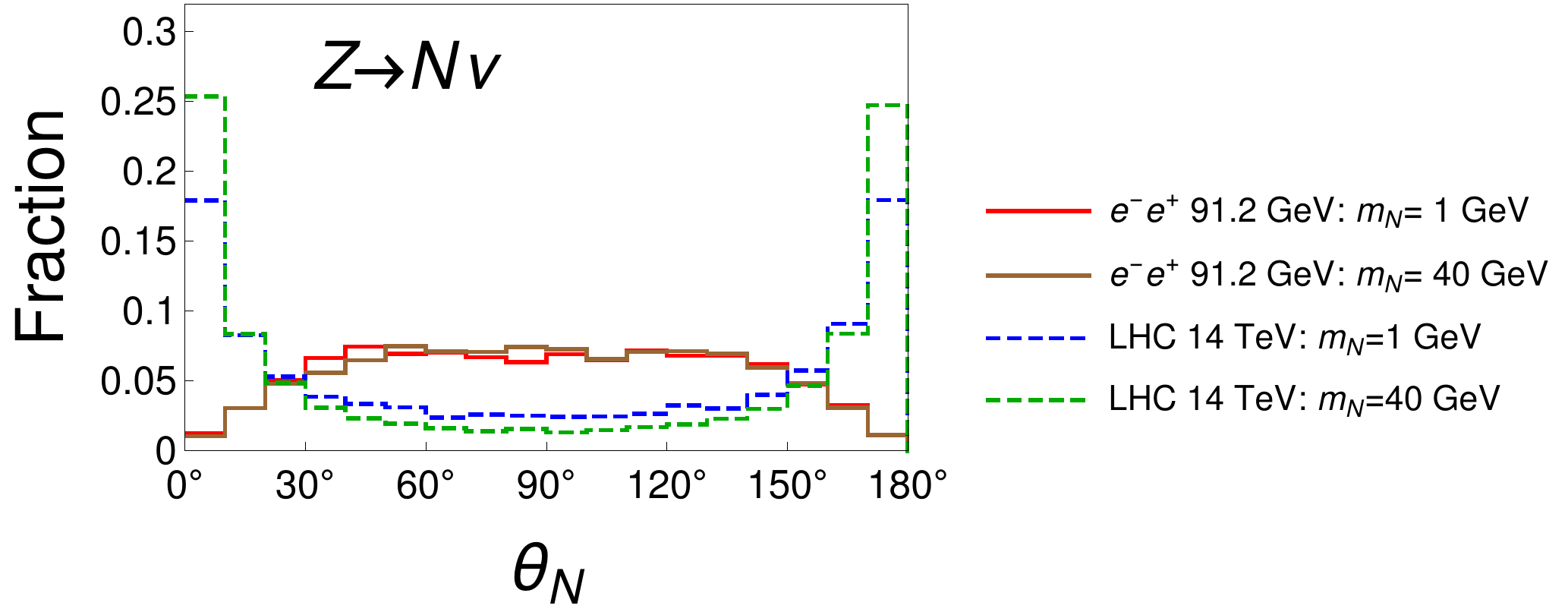}
	\par \vspace{0.3cm}
	\includegraphics[width=\columnwidth]{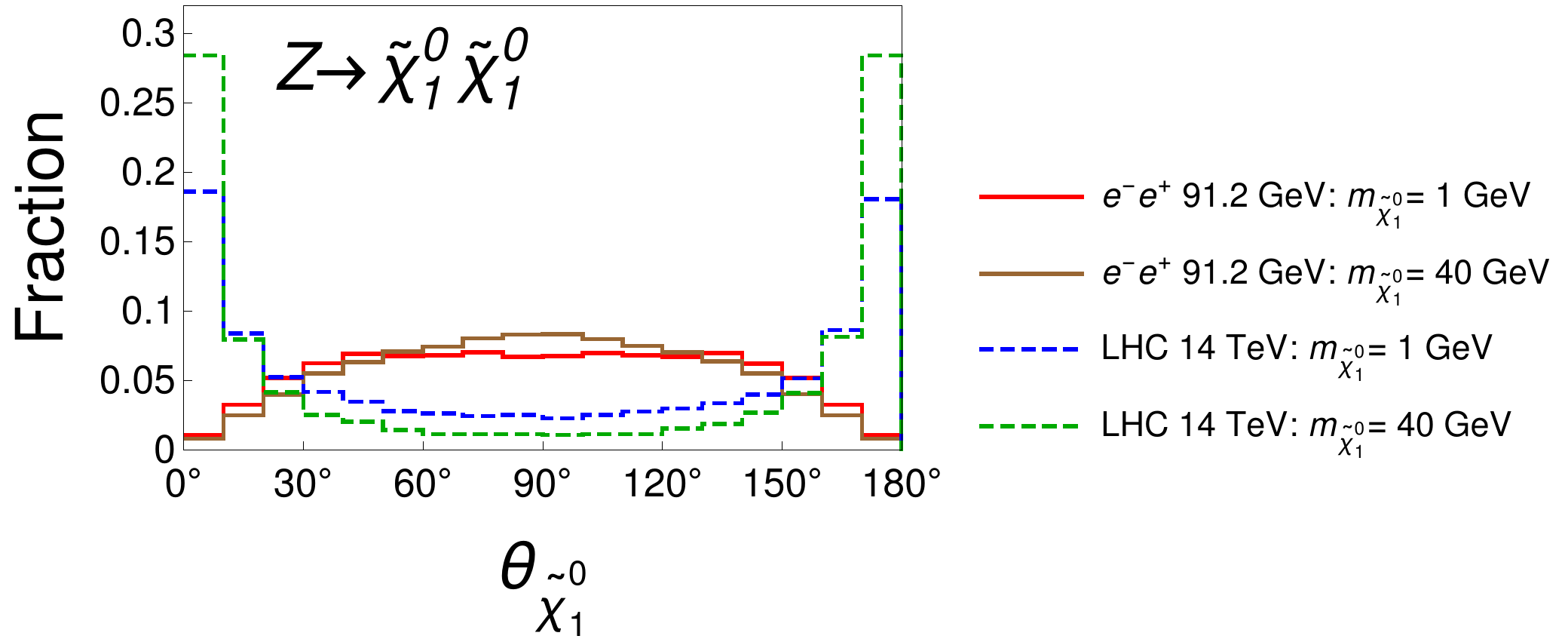}
	\caption{Probability distributions of the polar angle $\theta$ of the relevant LLPs in each physics scenario for both the LHC $pp$-collision with $\sqrt{s}=14$ TeV and the $e^- e^+$-collision at the Higgs factory and $Z-$pole running modes. The first plot is for $h\rightarrow XX$ with two benchmark values of $m_X$: 0.5 and 10 GeV, while the latter two plots are for $Z\rightarrow N \nu$ and $Z\rightarrow \tilde{\chi}_1^0\tilde{\chi}_1^0$, respectively, with two benchmark values of $m_N$ and $m_{\tilde{\chi}_1^0}$: 1 and 40 GeV.
	For $h\rightarrow XX$ ($Z\rightarrow \tilde{\chi}_1^0\tilde{\chi}_1^0$), the distributions include both LLPs from the same Higgs ($Z-$) boson decay.
	}
	\label{fig:ThetaDistribution}
\end{figure}

In Fig.~\ref{fig:ThetaDistribution}, we present the probability distributions of the LLP's polar angle $\theta$ for the three physics scenarios, obtained by Pythia. We choose a set of benchmark masses of the LLPs ($m_X=0.5$ and 10 GeV for $h\rightarrow XX$, and $m_{N}, m_{\tilde{\chi}_1^0}=$1 and 40 GeV for $Z\rightarrow N\nu$ and $Z\rightarrow \tilde{\chi}_1^0\tilde{\chi}_1^0$), and compare the distributions at the LHC ($pp-$collision with $\sqrt{s}=14$ TeV) and at a general $e^- e^+$ collider ($\sqrt{s}=240$, 91.2 GeV for the Higgs factory and the $Z-$pole running modes, respectively).
For the Higgs and $Z-$bosons simulations at the LHC, we use the same switches in Pythia 8 as those at the $e^- e^+$ colliders, and set proton-proton collisions at $\sqrt{s}=14$ TeV. With this choice, all the SM production modes of a SM Higgs boson at the LHC are considered. As for the $Z-$boson simulation at the LHC, $s-$channel production from a pair of quarks is simulated.  
The decay modes of the Higgs bosons and $Z-$bosons at the LHC are considered to be the same as those at the $e^- e^+$ colliders.
These plots clearly reflect that while at the LHC the LLPs are peaked with a large longitudinal boost, at the $e^-e^+$ colliders the LLPs are more prone to travel in the transverse direction.

We comment on the reason why we do not consider a detector locating downstream toward the beam axis with no radial displacement w.r.t. the IP. This would have the same spirit as FASER which has been approved for construction at the LHC.
At a proton-proton collider such as the LHC, because of the parton distributions inside the protons, the scattering products of the collision (and hence the LLP produced therefrom) can be largely boosted in the longitudinal direction, leading to excellent sensitivities on LLPs with FASER and AL3X (and also with MATHUSLA which has both radial and longitudinal displacement as our FD2 does).
However, at $e^- e^+$ colliders with two beams of equal energy, as electrons and positrons are point-like particles, we do not expect a longitudinal boost in general. The Higgs bosons are mainly produced from a $s-$channel Higgsstrahlung process, and the HNLs and the lightest neutralinos considered in this work are produced from $s-$channel on-shell $Z-$boson decays. These ensure that the $\theta-$distribution of the relevant LLPs is peaked around $\pi/2$, leading to very limited sensitivities for a far detector located at the forward direction in the same spirit of FASER or AL3X.

\subsection{Average Decay Probabilities}

\begin{table*}[]
	\centering
	\begin{tabular}{c|c|c|c|c|c|c|c}
	\hline
	\hline
	        \multirow{2}{*}{\,\,\,\,\,\,\,\,\,\,\,\,\,\,\,\,}
     	& \multirow{2}{*}{$\,\,\,D~[\text{m}]\,\,\,$} 
		&  \multirow{2}{*}{$\,\,\,\epsilon^{h\rightarrow XX}\cdot c\tau~[\text{m}]\,\,\,$} 
		&  \multirow{2}{*}{$\,\,\,\epsilon^{Z\rightarrow N\nu}\cdot c\tau~[\text{m}]\,\,\,$}
	    &  \multirow{2}{*}{$\,\,\,\epsilon^{Z\rightarrow \tilde{\chi}_1^0\tilde{\chi}_1^0}\cdot c\tau~[\text{m}]\,\,\,$} 
	    &  \multirow{2}{*}{$\,\,\,\frac{\epsilon^{h\rightarrow XX}}{\epsilon^{h\rightarrow XX}_{\text{CEPC}}}+1\,\,\,$} 
	    &  \multirow{2}{*}{$\,\,\,\frac{\epsilon^{Z\rightarrow N \nu}}{\epsilon^{Z\rightarrow N \nu}_{\text{CEPC}}}+1\,\,\,$}
	    &  \multirow{2}{*}{$\,\,\,\frac{\epsilon^{Z\rightarrow \tilde{\chi}^0_1\tilde{\chi}^0_1}}{\epsilon^{Z\rightarrow \tilde{\chi}^0_1\tilde{\chi}^0_1}_{\text{CEPC}}}+1\,\,\,$} \\
	    & & & & & & & \\ \hline
		\multicolumn{1}{c|}{\multirow{2}{*}{FD1}} & \multicolumn{1}{c|}{\,\,\,\,\,\,5}   & \multicolumn{1}{c|}{$4.7\times 10^{-2}$}                      & \multicolumn{1}{c|}{$6.7\times 10^{-2}$}                        & \multicolumn{1}{c|}{$6.6\times 10^{-2}$}                                                    & \multicolumn{1}{c|}{\,\,\,2.4}                                                           & \multicolumn{1}{c|}{\,\,\,2.4}                                                                 & \multicolumn{1}{c}{\,\,\,2.3}                                                                                                                       \\ \cline{2-8}
		\multicolumn{1}{c|}{}                    & \multicolumn{1}{c|}{\,\,\,10}   & \multicolumn{1}{c|}{$2.4\times 10^{-2}$}                      & \multicolumn{1}{c|}{$3.2\times 10^{-2}$}                        & \multicolumn{1}{c|}{$3.2\times 10^{-2}$}                                                    & \multicolumn{1}{c|}{\,\,\,1.7}                                                           & \multicolumn{1}{c|}{\,\,\,1.7}                                                                 & \multicolumn{1}{c}{\,\,\,1.7}                                                                                                                       \\ \hline
		\multicolumn{1}{c|}{\multirow{2}{*}{FD2}} & \multicolumn{1}{c|}{\,\,\,50}  & \multicolumn{1}{c|}{$3.5\times 10^{-2}$}                      & \multicolumn{1}{c|}{$6.3\times 10^{-2}$}                        & \multicolumn{1}{c|}{$6.3\times 10^{-2}$}                                                    & \multicolumn{1}{c|}{\,\,\,2.0}                                                           & \multicolumn{1}{c|}{\,\,\,2.3}                                                                 & \multicolumn{1}{c}{\,\,\,2.3}                                                                                                                          \\ \cline{2-8}
		\multicolumn{1}{c|}{}                    & \multicolumn{1}{c|}{100} & \multicolumn{1}{c|}{$1.9\times 10^{-2}$}                      & \multicolumn{1}{c|}{$3.3\times 10^{-2}$}                        & \multicolumn{1}{c|}{$3.3\times 10^{-2}$}                                                    & \multicolumn{1}{c|}{\,\,\,1.6}                                                           & \multicolumn{1}{c|}{\,\,\,1.7}                                                                 & \multicolumn{1}{c}{\,\,\,1.7}                                                                                                                       \\ \hline
		\multicolumn{1}{c|}{\multirow{2}{*}{FD3}} & \multicolumn{1}{c|}{\,\,\,50}  & \multicolumn{1}{c|}{$1.1\times 10^{-1}$}                      & \multicolumn{1}{c|}{$1.5\times 10^{-1}$}                        & \multicolumn{1}{c|}{$1.5\times 10^{-1}$}                                                    & \multicolumn{1}{c|}{\,\,\,4.3}                                                           & \multicolumn{1}{c|}{\,\,\,4.1}                                                                 & \multicolumn{1}{c}{\,\,\,4.1}                                                                                                                          \\ \cline{2-8}
		\multicolumn{1}{c|}{}                    & \multicolumn{1}{c|}{100} & \multicolumn{1}{c|}{$5.8\times 10^{-2}$}                      & \multicolumn{1}{c|}{$7.0\times 10^{-2}$}                        & \multicolumn{1}{c|}{$7.0\times 10^{-2}$}                                                    & \multicolumn{1}{c|}{\,\,\,2.7}                                                           & \multicolumn{1}{c|}{\,\,\,2.4}                                                                 & \multicolumn{1}{c}{\,\,\,2.4}                                                                                                                       \\ \hline
		\multicolumn{1}{c|}{\multirow{2}{*}{FD4}} & \multicolumn{1}{c|}{\,\,\,50}  & \multicolumn{1}{c|}{$1.7\times 10^{-1}$}                      & \multicolumn{1}{c|}{$1.9\times 10^{-1}$}                        & \multicolumn{1}{c|}{$1.9\times 10^{-1}$}                                                    & \multicolumn{1}{c|}{\,\,\,5.9}                                                           & \multicolumn{1}{c|}{\,\,\,4.9}                                                                 & \multicolumn{1}{c}{\,\,\,4.8}                                                                                                                           \\ \cline{2-8}
		\multicolumn{1}{c|}{}                    & \multicolumn{1}{c|}{100} & \multicolumn{1}{c|}{$6.5\times 10^{-2}$}                      & \multicolumn{1}{c|}{$7.1\times 10^{-2}$}                        & \multicolumn{1}{c|}{$7.1\times 10^{-2}$}                                                    & \multicolumn{1}{c|}{\,\,\,2.9}                                                           & \multicolumn{1}{c|}{\,\,\,2.5}                                                                 & \multicolumn{1}{c}{\,\,\,2.5}                                                                                                                       \\ \hline
		\multicolumn{1}{c|}{\multirow{2}{*}{FD5}} & \multicolumn{1}{c|}{\,\,\,50}  & \multicolumn{1}{c|}{$3.7\times 10^{-1}$}                      & \multicolumn{1}{c|}{$4.8\times 10^{-1}$}                        & \multicolumn{1}{c|}{$4.8\times 10^{-1}$}                                                    & \multicolumn{1}{c|}{12.2}                                                           & \multicolumn{1}{c|}{10.8}                                                                 & \multicolumn{1}{c}{10.8}                                                                                                                         \\ \cline{2-8}
		\multicolumn{1}{c|}{}                    & \multicolumn{1}{c|}{100} & \multicolumn{1}{c|}{$2.0\times 10^{-1}$}                      & \multicolumn{1}{c|}{$2.3\times 10^{-1}$}                        & \multicolumn{1}{c|}{$2.3\times 10^{-1}$}                                                    & \multicolumn{1}{c|}{\,\,\,6.9}                                                           & \multicolumn{1}{c|}{\,\,\,5.7}                                                                 & \multicolumn{1}{c}{\,\,\,5.7}                                                                                                                       \\ \hline
		\multicolumn{1}{c|}{\multirow{2}{*}{FD6}} & \multicolumn{1}{c|}{\,\,\,50}  & \multicolumn{1}{c|}{$8.2\times 10^{-1}$}                      & \multicolumn{1}{c|}{1.2}                        & \multicolumn{1}{c|}{1.2}                                                    & \multicolumn{1}{c|}{25.0}                                                           & \multicolumn{1}{c|}{26.2}                                                                 & \multicolumn{1}{c}{26.2}                                                                                                                         \\ \cline{2-8}
		\multicolumn{1}{c|}{}                    & \multicolumn{1}{c|}{100} & \multicolumn{1}{c|}{$7.1\times 10^{-1}$}                      & \multicolumn{1}{c|}{1.0}                        & \multicolumn{1}{c|}{1.0}                                                    & \multicolumn{1}{c|}{21.9}                                                           & \multicolumn{1}{c|}{22.3}                                                                 & \multicolumn{1}{c}{22.4}                                                                                                                       \\ \hline
		\multicolumn{1}{c|}{\multirow{2}{*}{FD7}} & \multicolumn{1}{c|}{\,\,\,50}  & \multicolumn{1}{c|}{$2.5\times 10^{-2}$}                      & \multicolumn{1}{c|}{$2.7\times 10^{-2}$}                        & \multicolumn{1}{c|}{$2.9\times 10^{-2}$}                                                    & \multicolumn{1}{c|}{\,\,\,1.7}                                                           & \multicolumn{1}{c|}{\,\,\,1.6}                                                                 & \multicolumn{1}{c}{\,\,\,1.6}                                                                                                                         \\ \cline{2-8}
		\multicolumn{1}{c|}{}                    & \multicolumn{1}{c|}{100} & \multicolumn{1}{c|}{$1.3\times 10^{-2}$}                      & \multicolumn{1}{c|}{$1.4\times 10^{-2}$}                        & \multicolumn{1}{c|}{$1.4\times 10^{-2}$}                                                    & \multicolumn{1}{c|}{\,\,\,1.4}                                                           & \multicolumn{1}{c|}{\,\,\,1.3}                                                                 & \multicolumn{1}{c}{\,\,\,1.3}                                                                                                                       \\ \hline
		\multicolumn{1}{c|}{\multirow{2}{*}{FD8}} & \multicolumn{1}{c|}{\,\,\,50}  & \multicolumn{1}{c|}{$3.0\times 10^{-2}$}                      & \multicolumn{1}{c|}{$4.6\times 10^{-2}$}                        & \multicolumn{1}{c|}{$4.8\times 10^{-2}$}                                                    & \multicolumn{1}{c|}{\,\,\,1.9}                                                           & \multicolumn{1}{c|}{\,\,\,1.9}                                                                 & \multicolumn{1}{c}{\,\,\,2.0}                                                                                                                        \\ \cline{2-8}
		\multicolumn{1}{c|}{}                    & \multicolumn{1}{c|}{100}   & $1.5\times 10^{-2}$   &  $2.3\times 10^{-2}$    &   $2.3\times 10^{-2}$     &     \,\,\,1.4      &     \,\,\,1.5  &   \,\,\,1.5 \\
	\hline
	\hline
	\end{tabular}
	\caption{List of the average decay probabilities times the proper decay length of each FD design for LLPs of mass 1 GeV in each physics scenario, when the boosted decay length is much larger than $D$. 
	The last three columns indicate the gain factor in the average decay probability, when an additional far detector is added to the CEPC's baseline near detector. The average decay probability for the CEPC's baseline detector $\epsilon_{\text{CEPC}}$ is calculated by the method detailed in Ref.~\cite{Wang:2019orr}. 
	}
	\label{tab:detectorsefficiencies}
\end{table*}

\begin{figure}[]
	\centering
	\includegraphics[width=\columnwidth]{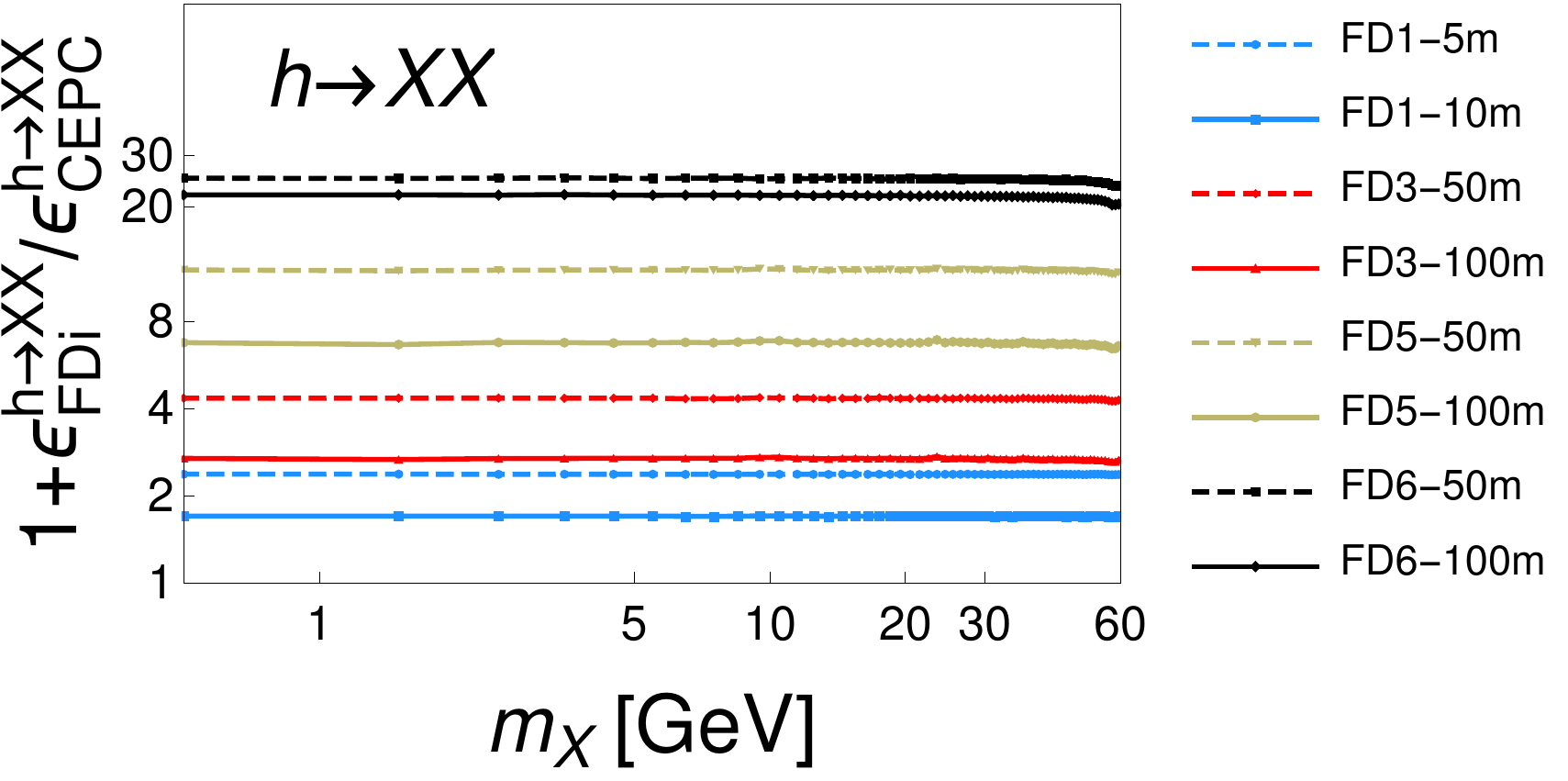}
	\par \vspace{0.3cm}
	\includegraphics[width=\columnwidth]{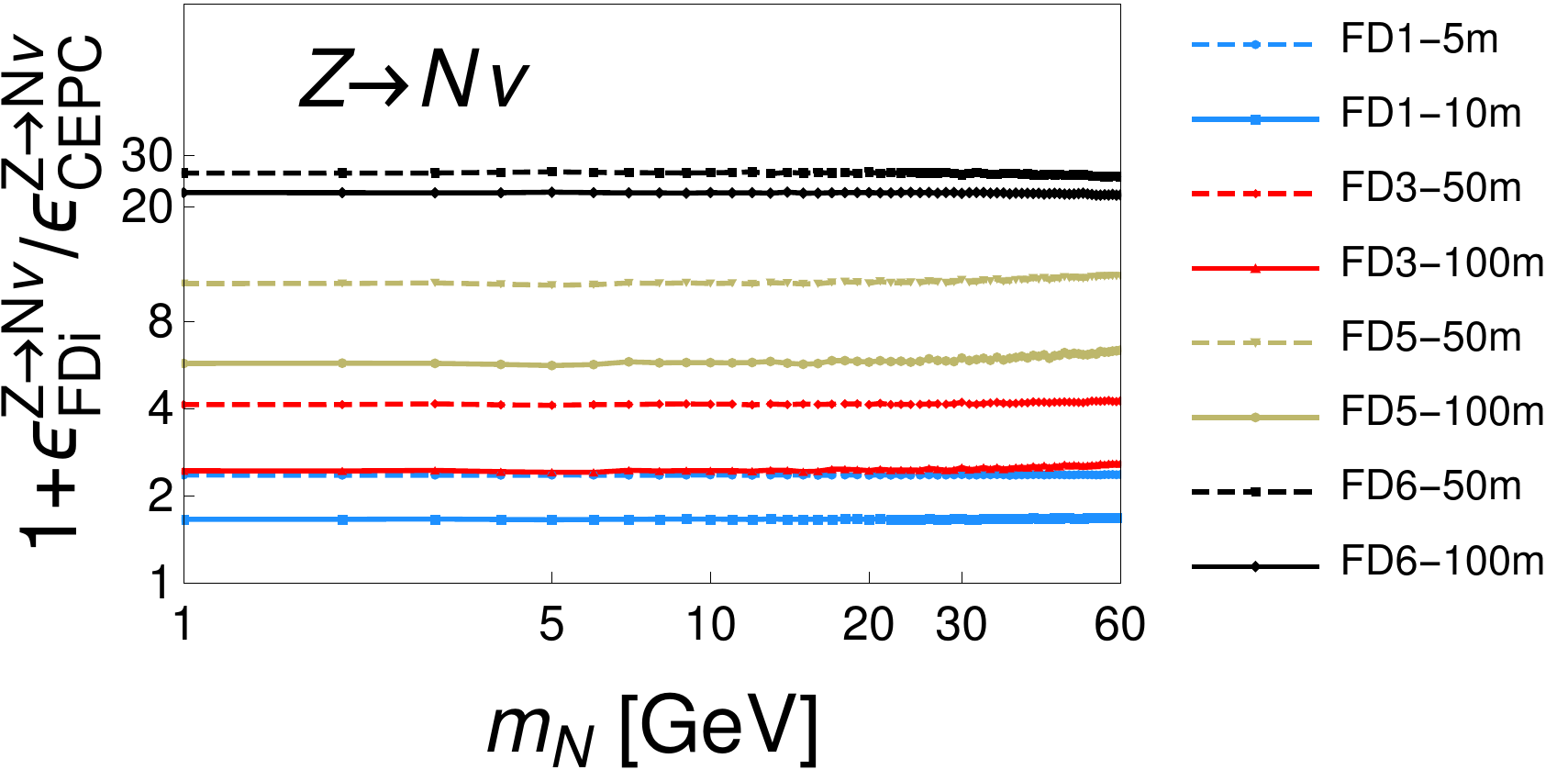}
	\par \vspace{0.3cm}
	\includegraphics[width=\columnwidth]{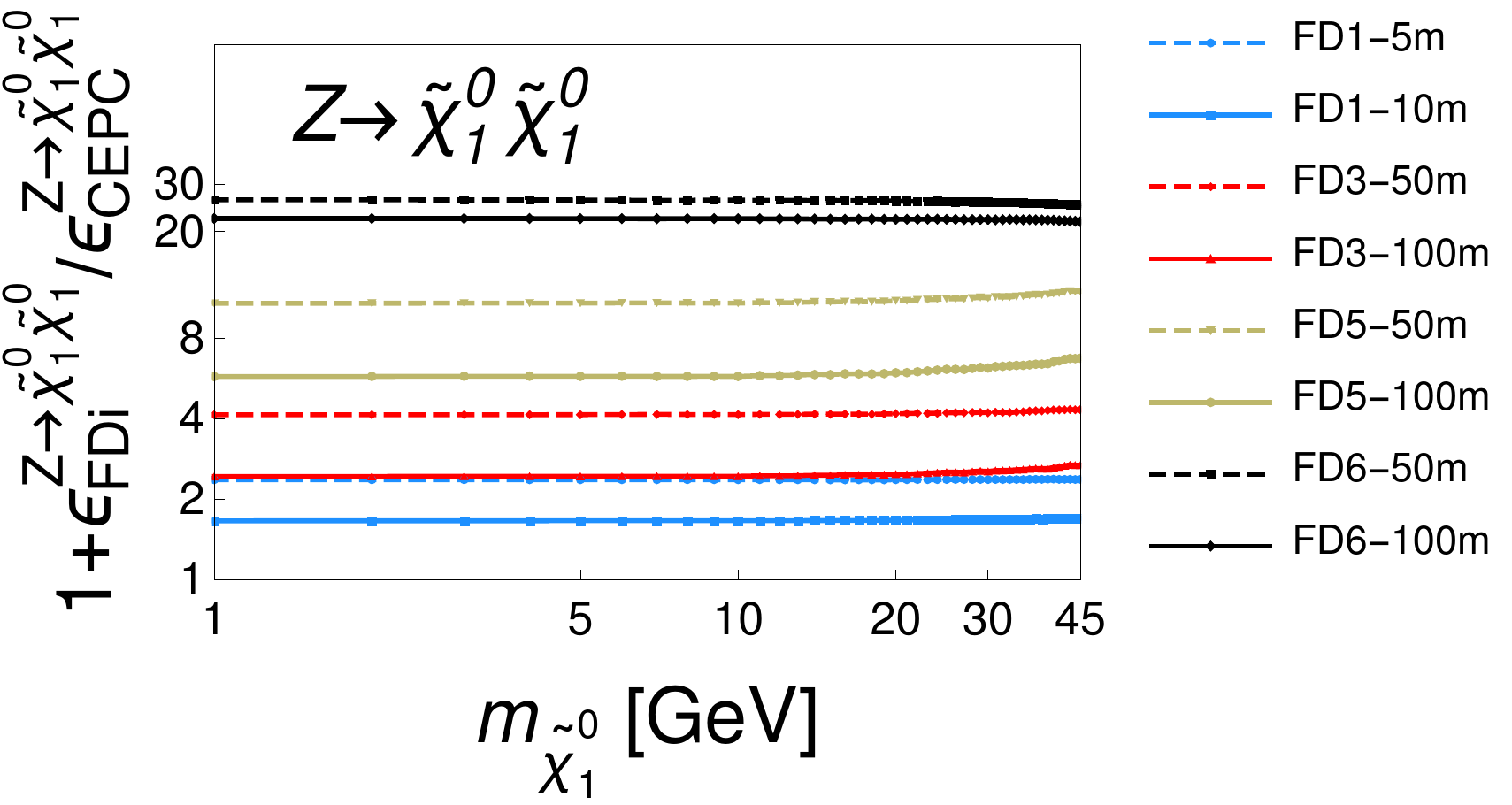}
	\caption{Average decay probability ratio as a function of the LLP mass in each physics study for FD1, FD3, FD5, and FD6. We vary the mass while keeping the proper decay length much larger than $\mathcal{O}(100\text{m})$, in order to obtain the average decay probability improvement with respect to the case of having only the CEPC's baseline detector. For each of FD1, FD3, FD5, and FD6, we show results of two choices of $D$.
	}
	\label{fig:FidEffRatio}
\end{figure}

Before we show and discuss the full numerical results, we present average decay probabilities $\epsilon \equiv \left\langle P[\text{LLP in f.v.}] \right\rangle$ of FD1$-$FD8 for each physics scenario with LLP's mass of 1 GeV. The average decay probability $\epsilon$ of a detector measures its acceptance of displaced vertices stemming from LLPs, and consequently comparing $\epsilon$ allows one to easily identify the optimal designs of potential far detectors at future $e^- e^+$ colliders, disregarding the impact of $N_h$ or $N_Z$.
We denote the average decay probability for each physic scenario as $\epsilon^{h\rightarrow XX}$, $\epsilon^{Z\rightarrow N \nu}$, and $\epsilon^{Z\rightarrow \tilde{\chi}_1^0\tilde{\chi}_1^0}$, respectively.
Since we are mainly interested in the long lifetime regime where the average decay probability is linearly dependent on $c\tau$, we present the results for $\epsilon\cdot c\tau$ in the limit with $\beta\gamma c\tau \gg D$.
This is done by requiring $c\tau \gsim \mathcal{O}(100)$ m, because at lepton colliders the mother particle $Z-$ or Higgs boson is produced approximately at rest in the lab frame, which renders $\beta\gamma \sim m_Z/(2$ GeV) or $m_h/(2$ GeV) for 1 GeV LLPs.

We list our results in Table~\ref{tab:detectorsefficiencies}, obtained by simulating one hundred thousand events. 
We find in general FD3-6 show the optimal results in the large decay length limit; this reflects the fact that the LLPs produced from $Z-$ and Higgs bosons decays travel transversely. 
FD1's average decay probabilities are weaker than FD3's by a factor of $\sim 1-2$; 
its close distance to the IP brings advantage of receiving more LLPs in its direction, but it is still beat by its disadvantage of much smaller volume than that of FD3.

Compared to FD3, FD2 is predicted with average decay probabilities worse by a factor of $\sim2-3$; its inferiority is due to its longitudinal displacement from the IP. 
Furthermore, we observe that FD4 has similar or slightly larger average decay probabilities than FD3. 
Compared to FD3, FD4 has a smaller base area covering fewer LLPs traveling in its windowed directions but it has a longer `fiducial path' for the LLPs traveling transversely. We conclude that these two effects more or less complement each other between FD3 and FD4. 

FD5 and FD6 clearly win over all other designs mainly by virtue of their much larger volumes. 
FD7 and FD8, as described previously, are shaped as a long rod of a base surface 20 m $\times$ 20 m and a length 2000 m, with FD7 placing along the $x-$direction and FD8 along the $z-$axis.  Table~\ref{tab:detectorsefficiencies} shows that in general they have weak potentials for LLP searches.

In order to find out the improvement on the average decay probabilities in the large $c\tau$ limit with and without a far detector, we present $(\epsilon/\epsilon_{\text{CEPC}}+1)$ in the last three columns of Table~\ref{tab:detectorsefficiencies}, where $\epsilon$ represents the average decay probability for a certain far detector and $\epsilon_{\text{CEPC}}$ denotes the average decay probability for the CEPC's baseline near detector calculated by the method detailed in Ref.~\cite{Wang:2019orr}.
The variable $(\epsilon/\epsilon_{\text{CEPC}}+1)$ thus is ratio of the average decay probability for the combination of both far and near detectors divided by that for the near detector only, which reflects the gain factor with an additional far detector.
We find that for FD3 and FD4 located 50 m from the IP, the gain factor is about 4-6, and for FD6, the factor can be as large as $\sim$25. 

In Fig.~\ref{fig:FidEffRatio}, we plot the $(\epsilon/\epsilon_{\text{CEPC}}+1)$ as a function of $m_{\text{LLP}}$ for each physics scenario, assuming the large $c\tau$ limit. We choose to show the curves of some representative designs only: FD1, FD3, FD5, and FD6 with both choices of $D$, because FD3 and FD4 have almost identical performance, and FD2, FD7, and FD8 have similar weak sensitivities.
In general, we observe that the enhancement in the average decay probabilities has a relatively small dependence on the LLP mass, and FD5 and FD6 clearly outperform FD1 and FD3 for all physics scenarios.

\section{Collider Sensitivities}
\label{sec:numericalresults}

In this section, we present numerical results for the three physics scenarios. 
We start with a discussion on possible background sources and shielding, and proceed to present sensitivity reaches of various detectors in the parameter space of each physics scenario.

For far-detector experiments at future lepton colliders, background sources for displaced-vertex searches mainly consist of cosmic rays, energetic muons from the IP, neutrino scatterings, and decays of the SM long-lived hadrons such as kaons and hyperons.
For FD2-FD8, when the far detector is located on the surface with a relatively large distance from the IP, there exist rock and concrete with depth of 50-100 meters between the far detector and the IP\footnote{For example, at the CEPC the main strata above the experiment hall can be full of rocks \cite{CEPCStudyGroup:2018rmc}.}.
Such shielding would already be sufficient to reducing the background of electrically charged particles and neutral hadrons from the main collisions, as shown in studies for MATHUSLA \cite{Chou:2016lxi,Curtin:2018mvb} and FASER \cite{Feng:2017uoz}.
As QCD activities at future lepton colliders running at the $Z$-pole and Higgs modes are expected to be lower than those at the LHC, similar amount and types of shielding should perform equally well.
The remaining background stems from cosmic rays, high-energy muons, and neutrino scattering, and can be essentially reduced by sufficient tracking on the charged particle direction of travel as well as more elaborate geometrical and timing cuts, following the discussion given in Refs.~\cite{Chou:2016lxi,Curtin:2018mvb}.
Since FD1 is close to the IP, different kinds of shielding would be required in order to suppress the background.
This can be inspired by the CODEX-b study \cite{Gligorov:2017nwh}, where it is shown that combination of 4.5 meters of lead or steel and 3-meter thick concrete shield wall should be sufficient to suppressing $K_L$, neutron and other hadronic backgrounds.
Besides, an active muon veto with an efficiency of $\mathcal{O}(10^{-5})$ can be embedded in the shield to reject backgrounds induced by muons or other charged particles.

Quantitative analysis on the background and shielding would require full Monte Carlo and detector simulation.
However, a realistic estimation of the background-rejection efficiencies relies on the detailed information of the detector performance.
Since the detector designs are just tentative proposals and the technologies are still under development, in order to simplify our analysis and focus on physics, in this study we assume $100\%$ detector efficiency and a background-free environment.
The same assumptions are made for the CEPC/FCC-ee near detectors.
We leave the detailed analyses with realistic estimates of all background processes and shielding for future studies.
We present the sensitivity results in terms of 3-signal-event contour curves which correspond to $95\%$ C.L. limits with zero background events.
The sensitivity limits for each physics scenario would be reduced to some extent according to the future realistic background studies.

\subsection{Exotic Higgs Decays}
\label{subsec:exotichiggsdecays}

\begin{figure}[]
	\centering
	\includegraphics[width=\columnwidth]{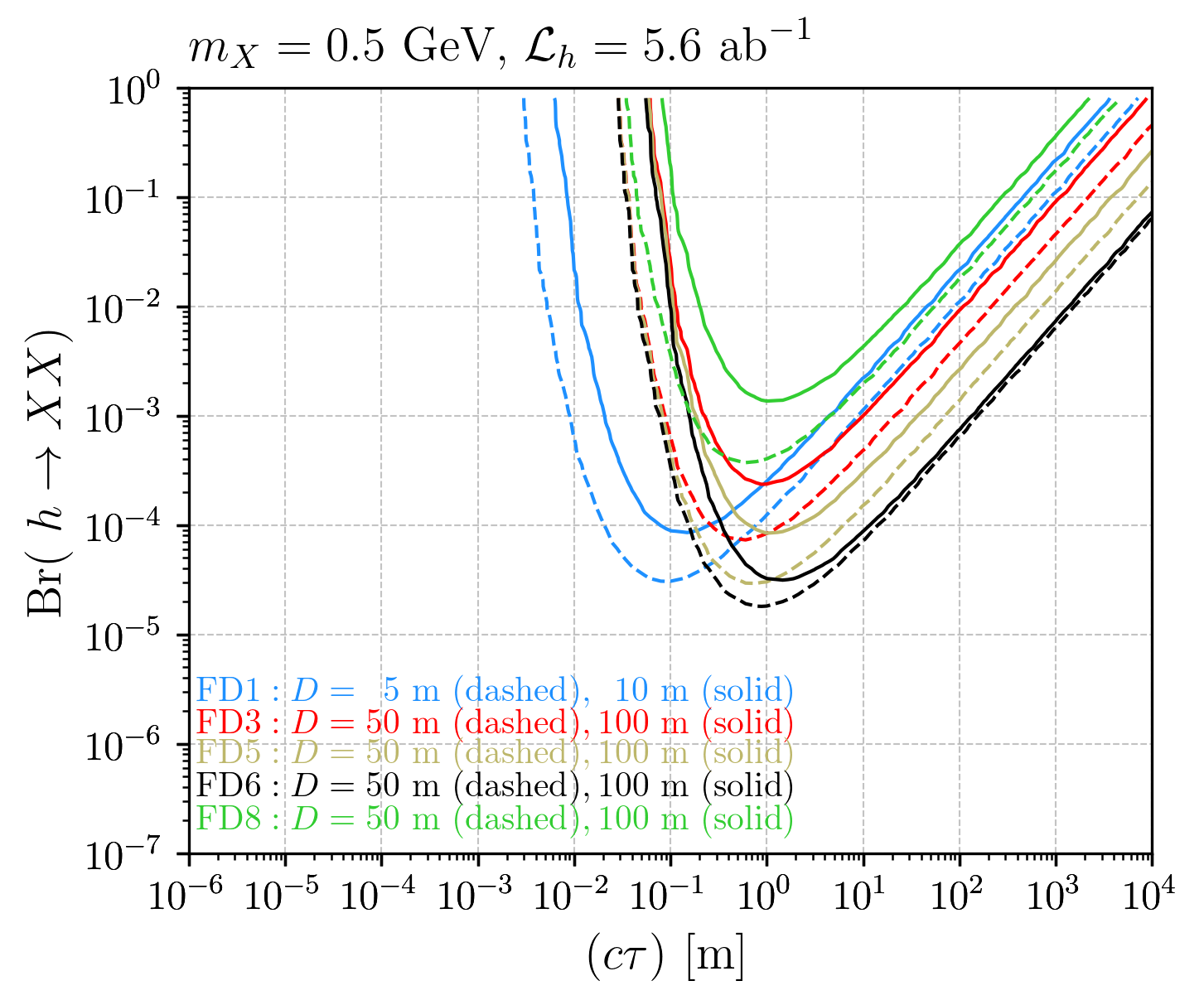}
	\includegraphics[width=\columnwidth]{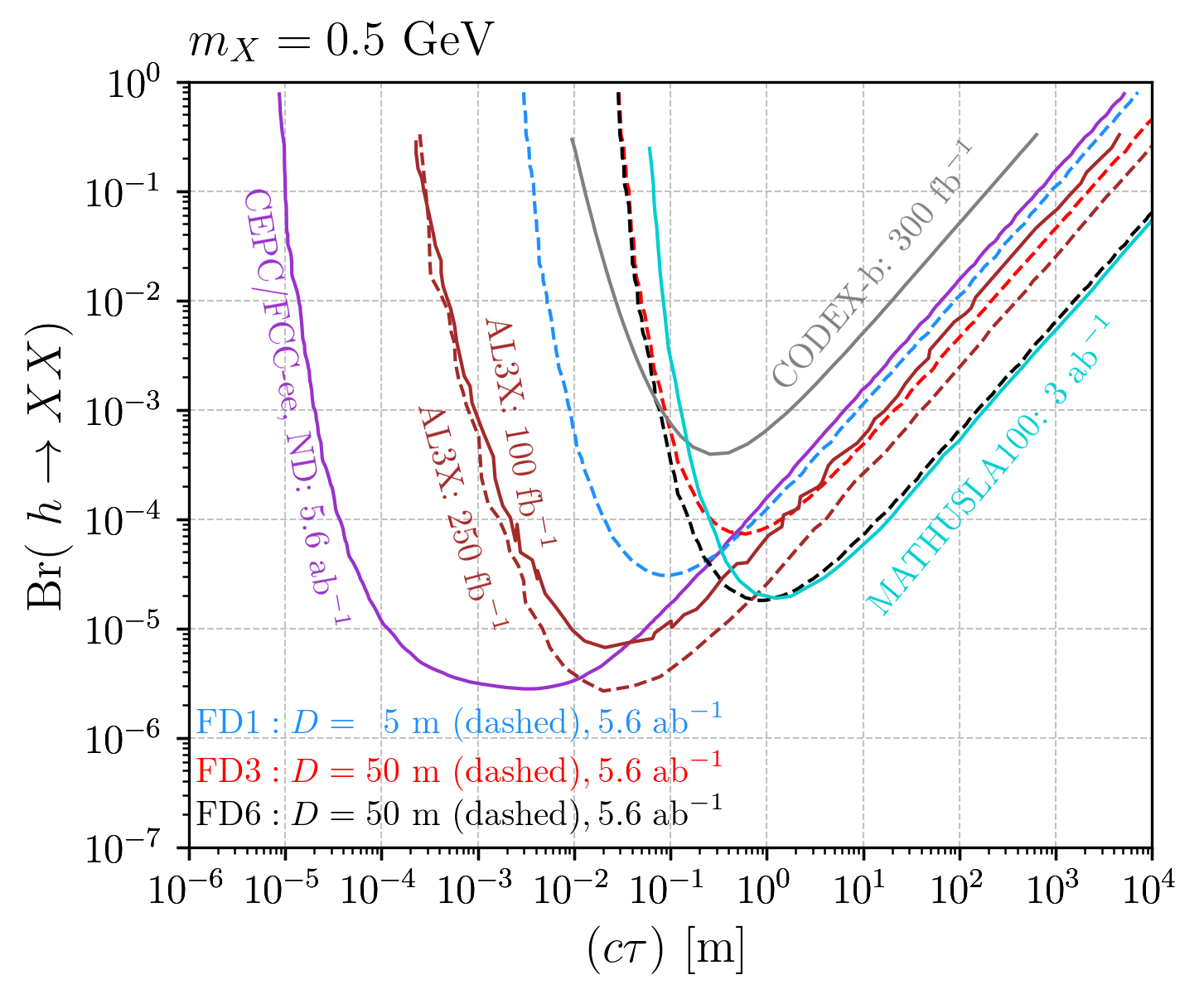}
	\caption{Upper: Sensitivity reaches of the CEPC/FCC-ee's far detectors FD1, FD3, FD5, FD6 and FD8 with different $D$ values in the Br($h \rightarrow XX$) vs. $c\tau$ plane for $m_X = 0.5$ GeV. 
	Lower: Sensitivity reaches of the CEPC/FCC-ee's far detectors FD1, FD3, FD6, compared with predictions for the CEPC/FCC-ee’s near detector (ND) and for AL3X, CODEX-b and MATHUSLA100.  
	 }
	\label{fig:H2XX-0p5}
\end{figure}
 
\begin{figure}[]
	\centering
	\includegraphics[width=\columnwidth]{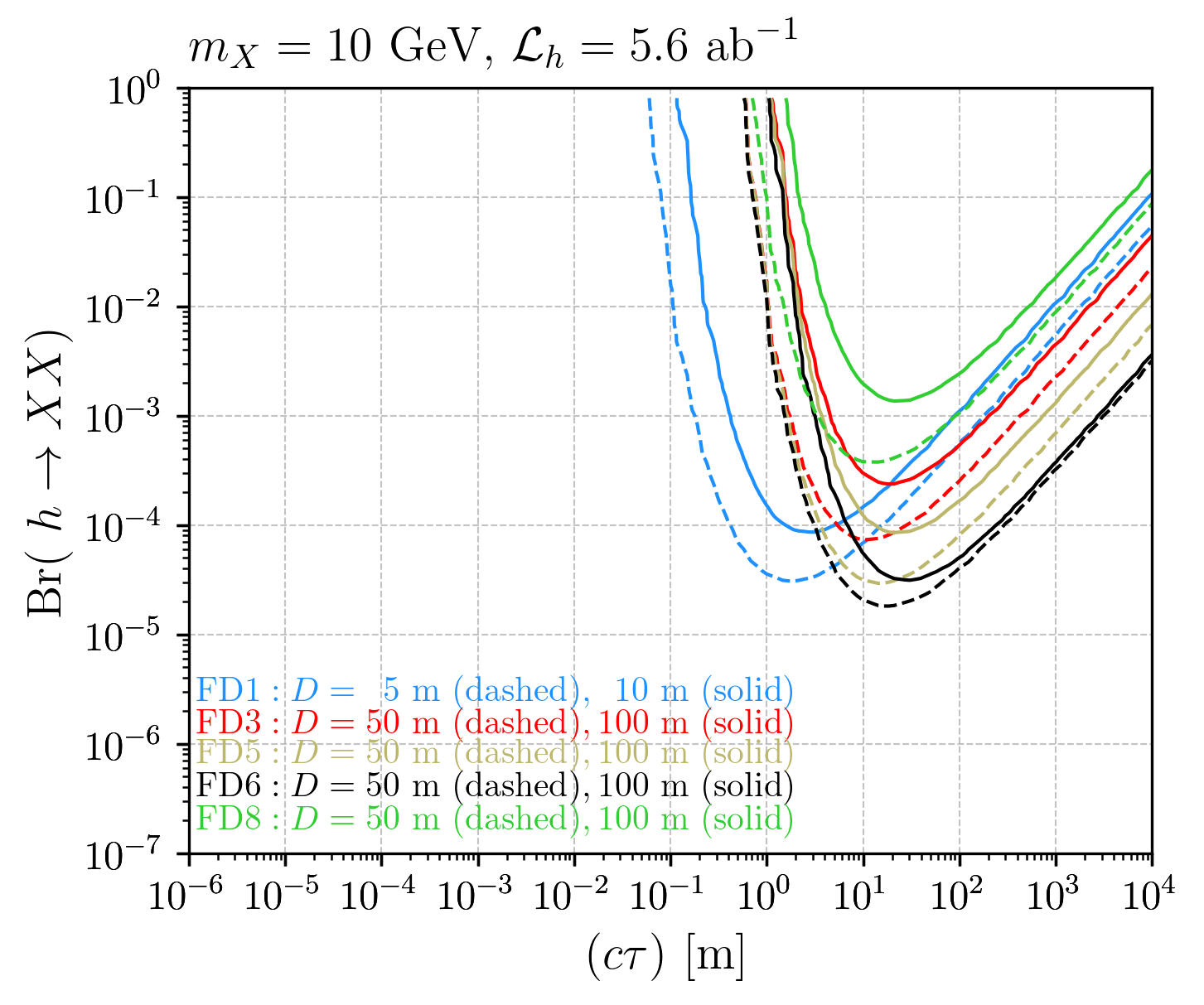}
	\includegraphics[width=\columnwidth]{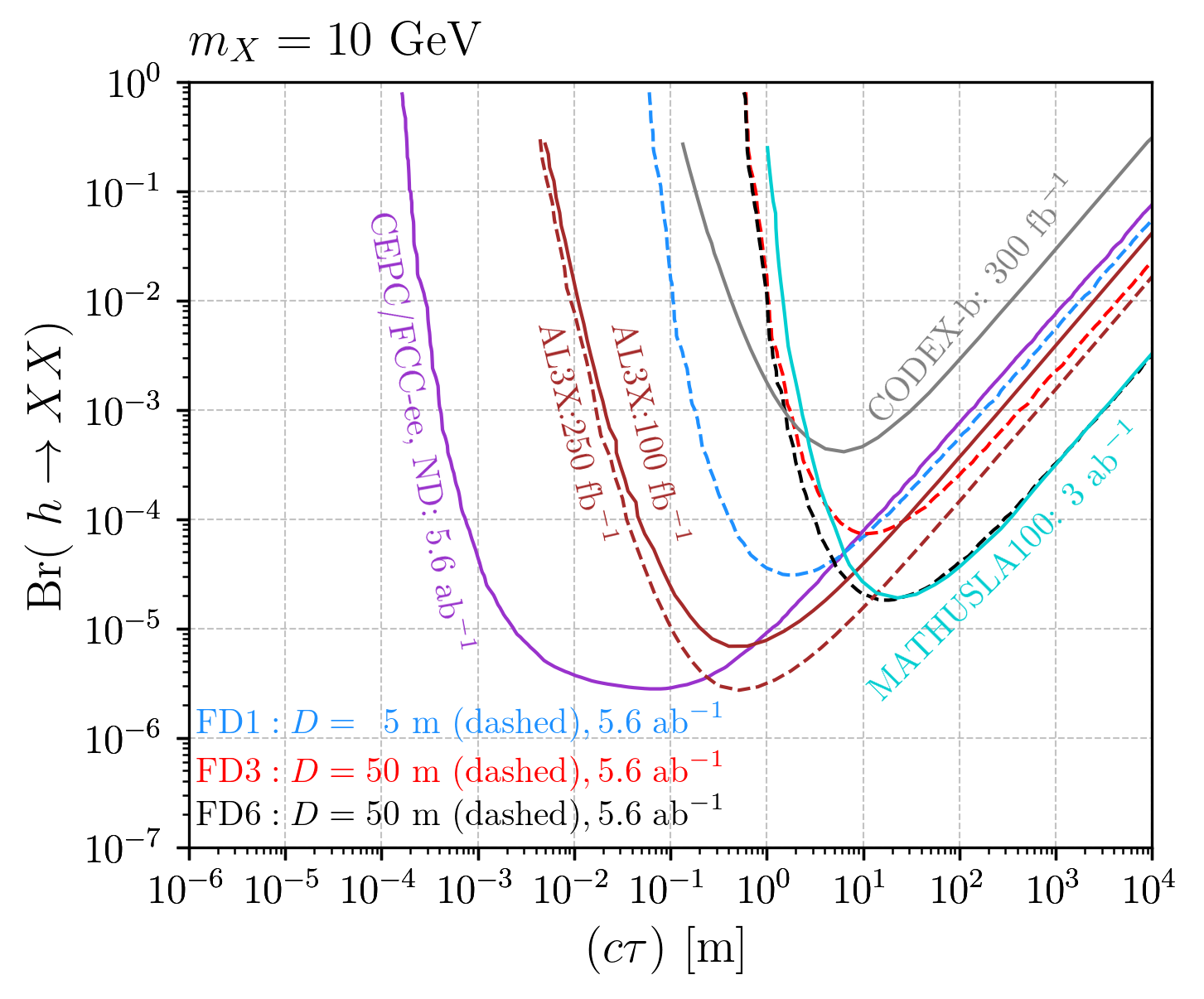}
	\caption{The same plots as Fig.~\ref{fig:H2XX-0p5} but for $m_X=10$ GeV.}
	\label{fig:H2XX-10}
\end{figure}

In Fig.~\ref{fig:H2XX-0p5} and Fig.~\ref{fig:H2XX-10}, we present the results in the Br$(h\rightarrow XX)$ vs. $c\tau$ plane for two benchmark values of $m_X=0.5$ and 10 GeV, respectively.
The $x-$axis label $c\tau$ is the proper decay length of the scalar particle $X$, while the $y-$axis label Br$(h\rightarrow XX)$ is the branching ratio of the Higgs boson decaying into a pair of $X$.
Each of Fig.~\ref{fig:H2XX-0p5} and Fig.~\ref{fig:H2XX-10} contains two plots. 
The upper plot shows the sensitivity reaches of FD1, FD3, FD5, FD6, and FD8 with both choices of $D$, and the lower plot compares the sensitivity projections of FD1, FD3, and FD6, together with those of the CEPC/FCC-ee's near detector, and of other future detectors at the LHC such as CODEX-b \cite{Gligorov:2017nwh}, MATHUSLA \cite{Alpigiani:2018fgd} and AL3X \cite{Gligorov:2018vkc}.
The limits are shown together in one plot to compare the discovery potentials of all different detectors.

As given in Table~\ref{tab:physicscases}, the CEPC and FCC-ee are estimated to produce almost the same number of Higgs bosons at $\sqrt{s}=240$ GeV ($N_h=1.14\times 10^6$ with $\mathcal{L}_h=5.6$ ab$^{-1}$).
Their near detectors are also very similar to each other, leading to the same sensitivity reaches shown in Fig.~\ref{fig:H2XX-0p5} and Fig.~\ref{fig:H2XX-10}, obtained with the formulas given in Ref.~\cite{Wang:2019orr}. 
The main difference in the sensitivity reaches between $m_X = 0.5$ and 10 GeV can be understood as a horizontal shift along the $x-$axis. 
This reflects the fact that when $m_X\ll m_h$, a change in $m_X$ leads primarily to a different distribution of $\beta\gamma$ of $X$. 

For both $m_X = $ 0.5 and 10 GeV, FD1(FD3) may reach approximately $3(6)\times 10^{-5}$ in Br($h\rightarrow XX$) while FD6 is expected to behave similarly as MATHUSLA100 (100 m $\times$ 100 m $\times$ 20 m) does at the LHC, reaching $\sim 2\times 10^{-5}$ in Br($h\rightarrow XX$). FD5 is expected to have a sensitivity strength between those of FD3 and FD6, while FD8 is predicted to have the weakest sensitivity. 
Also, as expected, for all designs, larger $D$ gives weaker reaches in Br($h\rightarrow XX$). Increasing $D$ from 50 m to 100 m for these designs reduces the Br($h\rightarrow XX$) reaches by a factor $\sim 2-5$. 

Compared with the proposed future experiments at the LHC, the far detectors at $e^- e^+$ colliders do not immediately have an advantage. This is mainly owing to the orders of magnitude difference in Higgs production between the high-luminosity LHC (HL-LHC) and the CEPC/FCC-ee. At the HL-LHC, with the projected integrated luminosity of 3 ab$^{-1}$, a total number of 1.8$\times 10^8$ Higgs bosons could be produced, which is over 150 times more than the number at the CEPC/FCC-ee. However, compared to the CEPC/FCC-ee's near detector, the far detector would have better sensitivities in the larger $c\tau$ region.

We comment that the most sensitive value of the proper decay length $c\tau$ of a far detector is determined from the boost factor of the LLPs and, roughly speaking, the distance $d$ from the IP to the mid-point of the detector, obeying the relation $c\tau \left< \beta \gamma \right>\sim d$ with $\left< \beta \gamma \right>$ denoting the average boost factor of the LLPs that travel inside the far detector's window.

\subsection{Heavy Neutral Leptons}
\label{subsec:HNL}

\begin{figure}[]
	\centering
	\includegraphics[width=\columnwidth, height=6.5cm]{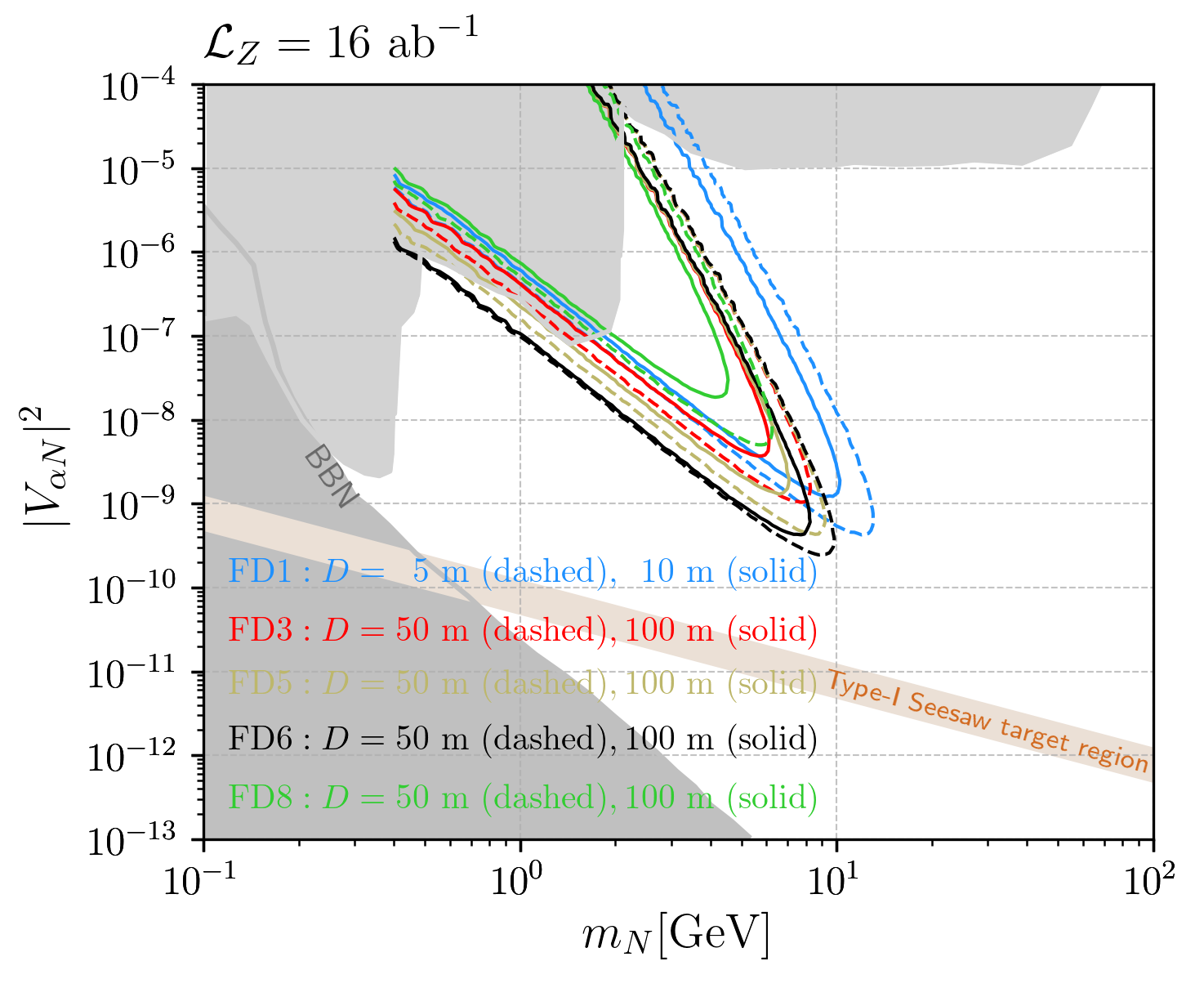}
	\includegraphics[width=\columnwidth, height=6.5cm]{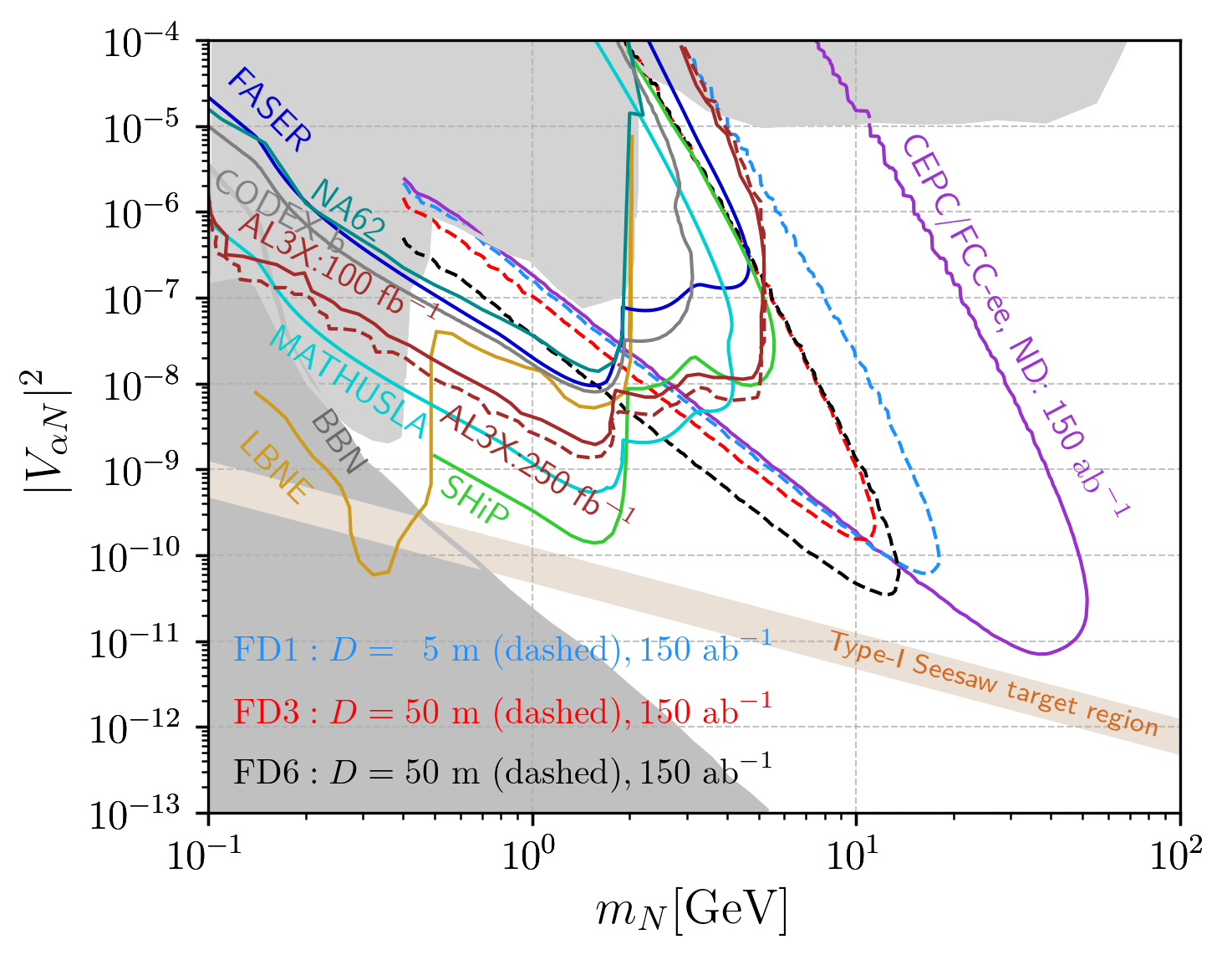}
	\includegraphics[width=\columnwidth, height=6.5cm]{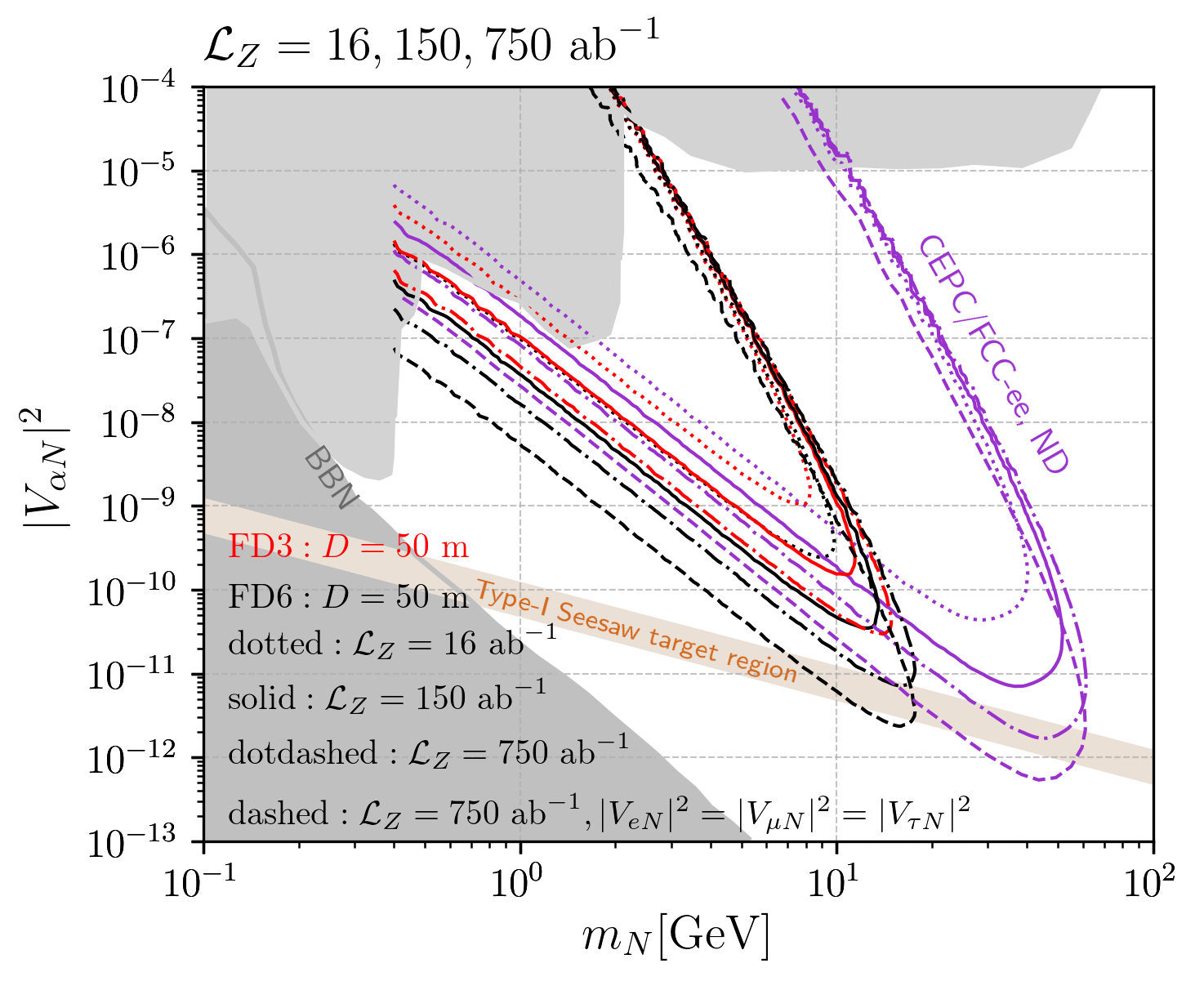}
	\caption{Upper: Sensitivity reaches of the CEPC/FCC-ee's far detectors with different $D$ values in the $|V_{\alpha N}|^2$ vs. $m_N$ plane. 
	Middle: Sensitivity reaches of the CEPC/FCC-ee's far detectors FD1, FD3, FD6, compared with predictions for the CEPC/FCC-ee's near detector (ND) and other experiments.
	Lower: Sensitivity reaches of both the far and near detectors at the CEPC/FCC-ee with three different integrated luminosities $\mathcal{L}_Z$.
}
\label{fig:HNL}
\end{figure}

We present our results in Fig.~\ref{fig:HNL}, showing the sensitivities in the $|V_{\alpha N}|^2$ ($\alpha=e / \mu$) vs. $m_N$ plane. In all the plots, the light gray shaded area in the upper area represents the experimentally excluded parameter space, given by combining  the search results of PS191 \cite{Bernardi:1987ek}, JINR \cite{Baranov:1992vq}, CHARM \cite{Bergsma:1985is} and DELPHI \cite{Abreu:1996pa} (see Ref.~\cite{Deppisch:2015qwa} for a review).
The brown band represents the ``Type-I Seesaw target region'' for $0.05$ eV $\lsim m_{\nu_\alpha} \lsim$ 0.12 eV with the relation $|V_{\alpha N}|^2\sim m_{\nu_\alpha} / m_N$, where $m_{\nu_\alpha}$ denotes the mass of the active neutrino $\nu_\alpha$ of generation $\alpha$. The lower limit 0.05 eV stems from the neutrino oscillation observation \cite{Canetti:2010aw} that there is at least one mass eigenstate of the active neutrinos of mass at least 0.05 eV. On the other hand, the upper limit comes from cosmological observation \cite{Aghanim:2018eyx} that the sum of the neutrino masses should be smaller than 0.12 eV.
The success of Big Bang Nucleosynthesis (BBN) leads also to bounds on the lifetime of the HNLs \cite{Ruchayskiy:2012si}. The parameter region disfavored by the primordial Helium$-4$ abundance is shown in dark gray.

In the upper plot of Fig.~\ref{fig:HNL}, we compare the sensitivity reaches of the representative far detectors FD1, FD3, FD5, FD6, and FD8 for both options of $D$, assuming the integrated luminosity $\mathcal{L}_Z$ of a general $e^- e^+$ collider running at the $Z-$pole is 16 ab$^{-1}$.
We observe that FD1 has the largest mass reach by virtue of its closeness to the IP, while FD3 and FD6 may reach $m_N\sim 7-9$ GeV with $|V_{\alpha N}|^2 = 1 \times 10^{-9}$ and $2 \times 10^{-10}$, respectively.
FD5's sensitivity is between those of FD3 and FD6. 
FD8 is projected with the weakest limits. 

The middle plot compares the sensitivities of CEPC/FCC-ee's far detectors FD1, FD3 and FD6 with those of LHC experiments and of near detectors at the CEPC/FCC-ee. 
We extract the sensitivity reaches of CODEX-b (300 fb$^{-1}$), FASER (3 ab$^{-1}$) and MATHUSLA (3 ab$^{-1}$) from Ref.~\cite{Helo:2018qej}, AL3X (100 and 250 fb$^{-1}$) from Ref.~\cite{Dercks:2018wum}, SHiP ($2\times 10^{20}$ pot) from Ref.~\cite{Bondarenko:2018ptm}, LBNE from Ref.~\cite{Adams:2013qkq}, and NA62 from Ref.~\cite{Drewes:2018gkc}.
The projections for the near detectors at the CEPC/FCC-ee are estimated according to the calculation procedure given in Ref.~\cite{Wang:2019orr}. 
We find that the far detectors with $\mathcal{L}_Z=150$ ab$^{-1}$ are sensitive to $m_N$ in the GeV range in comparison with the sensitive range $\mathcal{O}(10)$ GeV of the near detectors at the CEPC/FCC-ee. The reason for this difference is that HNLs with smaller masses have a longer lifetime and decay length, rendering them prone to decay inside the far detector. 
The near detector at the CEPC/FCC-ee covers almost all sensitive regions of FD1, but FD3 and FD6 may probe smaller $|V_{\alpha N}|^2$ than the near detector for $m_N\lsim10$ GeV.

Compared to the future far detectors at the LHC, FD1, FD3 and FD6 clearly explore more regions in the parameter space corresponding to larger $m_N$ and smaller $|V_{\alpha N}|^2$.
The reason is as follows.
For the FDs at the LHC, the dominant contributions to long-lived HNLs come from $D-$ and $B-$mesons decays while the $Z-$boson decays give only peripheral contributions because of its smaller production cross section and lower acceptance at these FDs.
In this study, we consider HNLs produced from $Z-$decays for FDs at the lepton colliders.
Such HNLs have a larger kinematically allowed mass range, compared to those produced from mesons decays.
Also, the FDs at the lepton colliders have a better acceptance for the HNLs produced from $Z-$boson decays, especially for HNLs heavier than $B-$mesons, mainly as a result of the position of the far detectors and the kinematics of the HNLs.

In the lower plot we compare the performance of the CEPC/FCC-ee ND, FD3, and FD6 for a variety of integrated luminosities: $\mathcal{L}_Z=\mathcal{L}_Z^{\text{CEPC}}$, $\mathcal{L}_Z^{\text{FCC-ee}}$,  and $5\, \mathcal{L}_Z^{\text{FCC-ee}}$, where $\mathcal{L}_Z^{\text{CEPC}}=16$ ab$^{-1}$, $\mathcal{L}_Z^{\text{FCC-ee}}=150$ ab$^{-1}$~\cite{Abada:2019zxq}, and $5\, \mathcal{L}_Z^{\text{FCC-ee}}$ would correspond to roughly 10-year running at the $Z-$pole for the current FCC-ee design with four IPs.
Projections are shown for both the near and far detectors at the CEPC/FCC-ee.
For $\mathcal{L}_Z=750$ ab$^{-1}$, we find that FD6 may reach $\sim 10^{-11}$ for $m_N$ between 10 and 20 GeV.
Furthermore, the previous limits all assume only one single HNL mixes with one single generation of active neutrino generations.
If one HNL has equal mixings with all three active neutrino generations, i.e. $|V_{e N}|^2=|V_{\mu N}|^2=|V_{\tau N}|^2$, we find that with $\mathcal{L}_Z=750$ ab$^{-1}$, the combination of FD6 and the near detector at the CEPC or FCC-ee may probe the Type-I Seesaw limits on $|V_{\alpha N}|^2$ for $m_N$ between 10 and 60 GeV, if $m_{\nu_\alpha}$ lies within the considered range.

\subsection{Light Neutralinos from $Z-$boson Decays}
\label{subsec:lightneutralinos}

\begin{figure}[]
	\centering
	\includegraphics[width=\columnwidth, height=6.5cm]{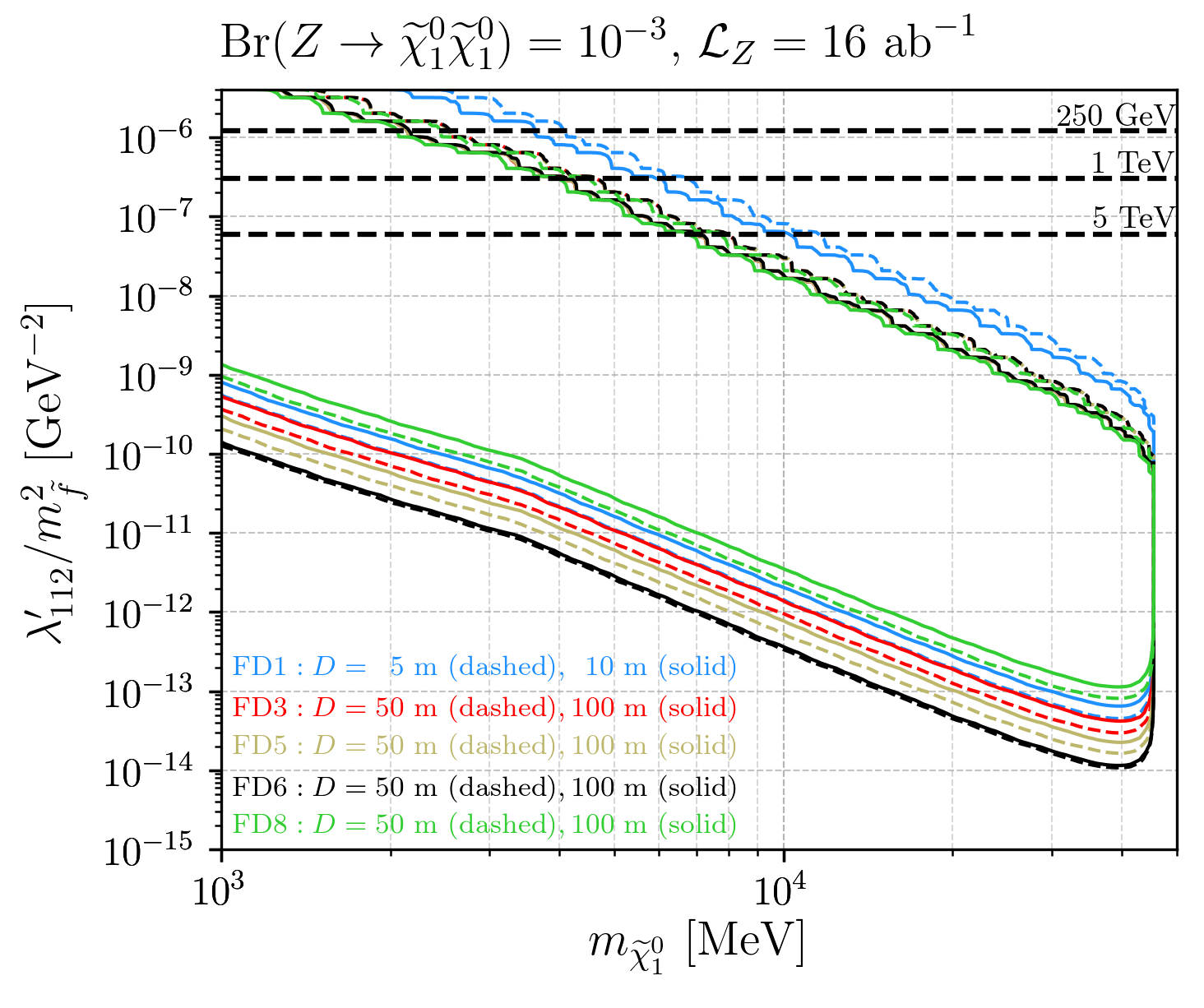}
	\includegraphics[width=\columnwidth, height=6.5cm]{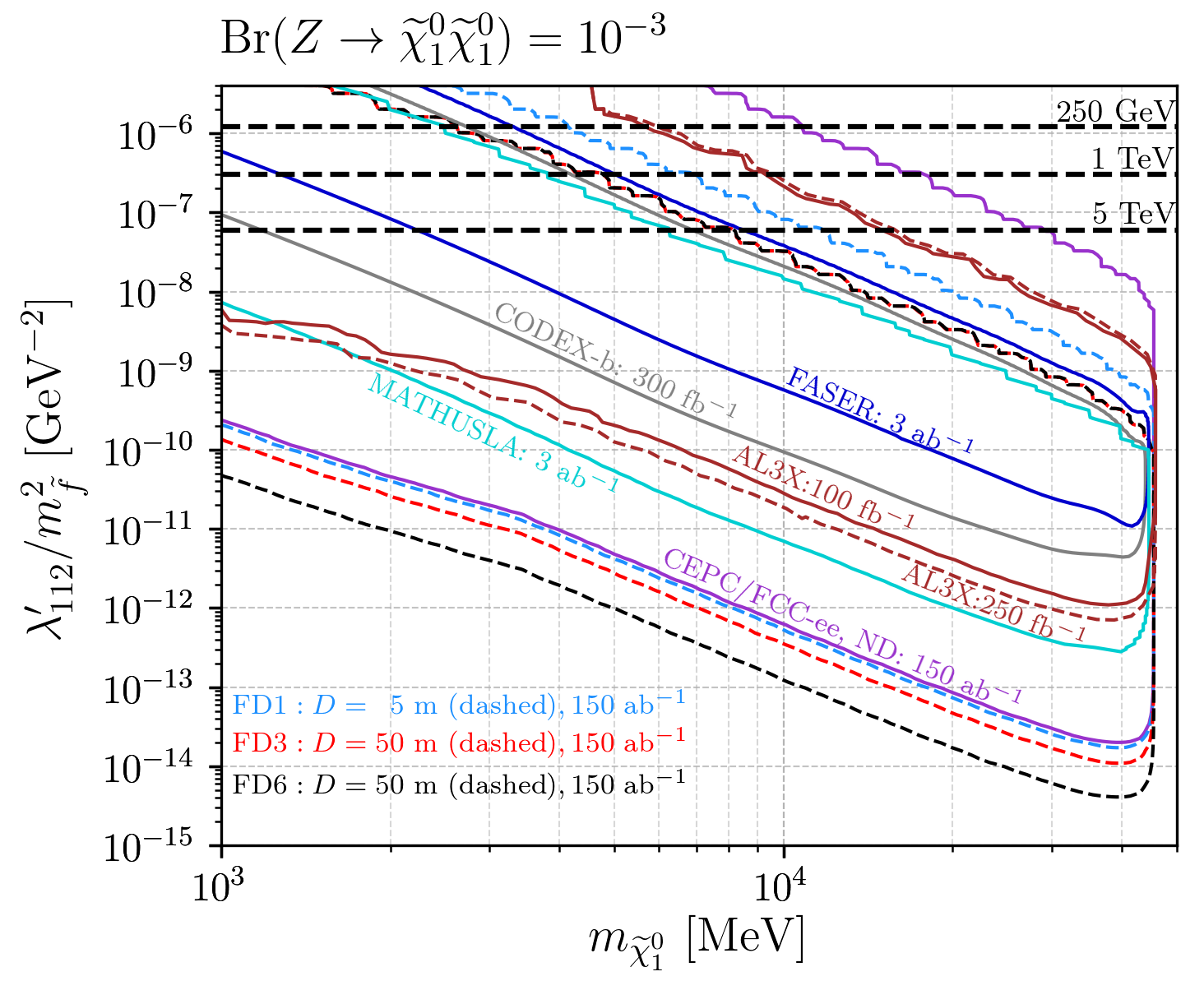}
	\includegraphics[width=\columnwidth, height=6.5cm]{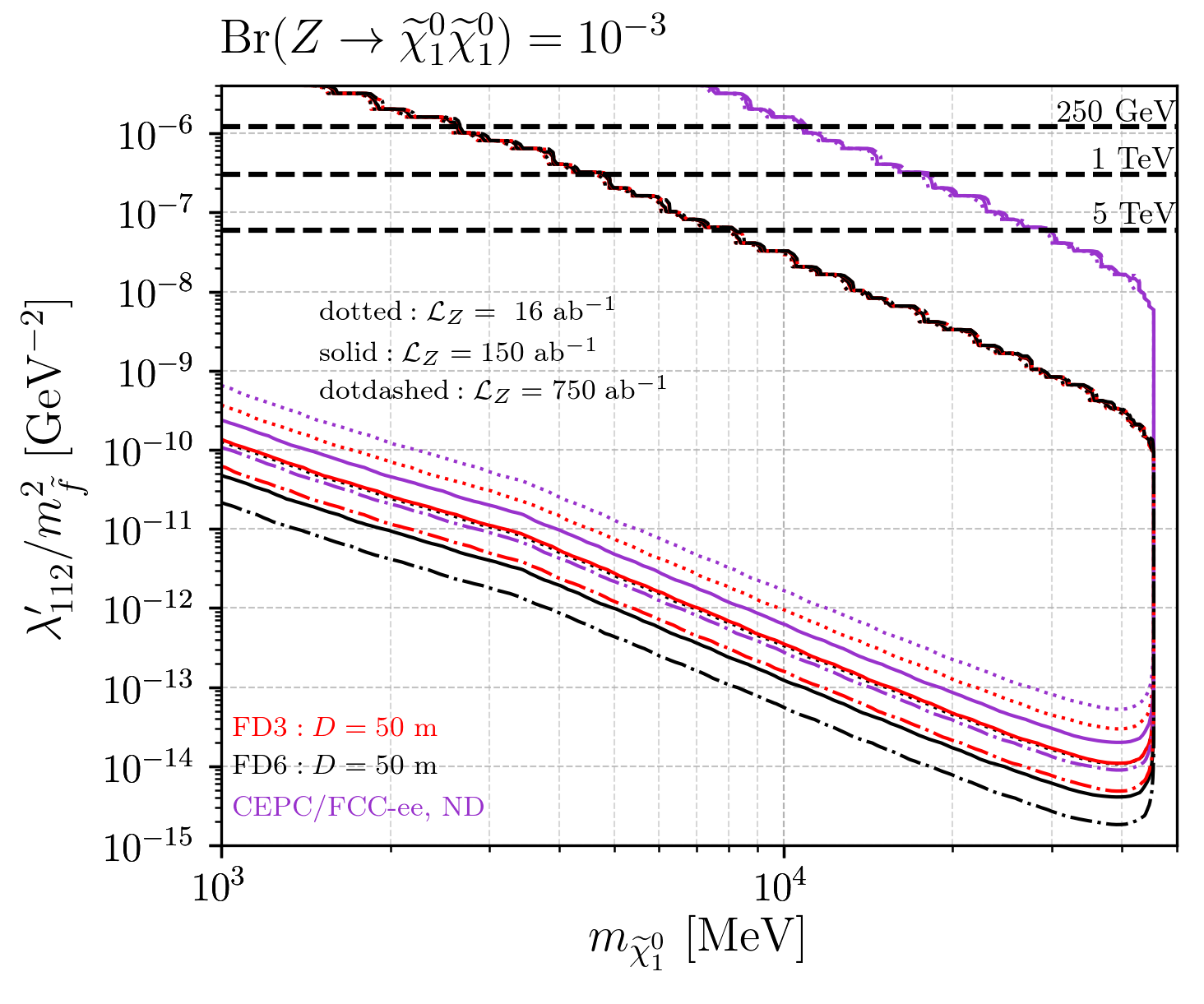}
	\caption{Sensitivity reaches of different experiments assuming Br($Z\rightarrow\tilde{\chi}_1^0\tilde{\chi}_1^0$) = $10^{-3}$.  Upper: Limits of the CEPC/FCC-ee's far detectors with different $D$ values in the $\lambda'_{112}/m^2_{\tilde{f}}$ vs. $m_{\tilde{\chi}_1^0}$ plane, 
	Middle: Limits of the CEPC/FCC-ee's far detectors FD1, FD3, FD6, compared with predictions for the CEPC/FCC-ee's near detector (ND) and other experiments.
	Lower: Limits of both the far and near detectors at the CEPC/FCC-ee with different integrated luminosities $\mathcal{L}_Z$.
	}
	\label{fig:Z2n1n1:BrLim}
\end{figure}

In Fig.~\ref{fig:Z2n1n1:BrLim}, 
the upper plot shows the sensitivity reach of FD1, FD3, FD5, FD6, and FD8 with both values of $D$ at a future $e^-e^+$ collider.
The middle plot compares the sensitivity reaches of representative far detectors with those of the CEPC/FCC-ee's near detector, and future experiments at the LHC.
As in the HNL case, the integrated luminosities $\mathcal{L}_Z$ in the upper plot are chosen to be $\mathcal{L}_Z=\mathcal{L}_Z^{\text{CEPC}}$ , while those in the middle plot for the FDs and ND at lepton collider are $\mathcal{L}_Z=\mathcal{L}_Z^{\text{FCC-ee}}$.
The lower plot presents the potential reaches of FD3, FD6, and a near detector at the CEPC/FCC-ee with $\mathcal{L}_Z=16$ ab$^{-1}$, 150 ab$^{-1}$ and 750 ab$^{-1}$ integrated luminosities.

In each plot, the black dashed horizontal lines correspond to the latest upper bound on the single RPV coupling $\lambda'_{112}$ assuming different sfermion mass values of 250 GeV, 1 TeV and 5 TeV \cite{Kao:2009fg}, given by:
\begin{eqnarray}
\lambda'_{112} < 0.6 \times \frac{m_{\tilde{s}_R}}{2\text{ TeV}}.
\end{eqnarray}
The 3-signal-event isocurves of the near detectors at the CEPC/FCC-ee are reproduced from Ref.~\cite{Wang:2019orr} by adopting  the CEPC's baseline detector, and the predictions for future LHC detectors (CODEX-b, FASER, MATHUSLA and AL3X) are extracted from Refs.~\cite{Helo:2018qej,Dercks:2018wum} for the same physics scenario. 

All detectors have a mass reach from $\sim 1$ GeV to $\sim m_Z/2$. 
Among FD1, FD3, and FD6, we find that FD6 could probe lower $\lambda'_{112}/m^2_{\tilde{f}}$ and hence outperforms the others. All of FD1, FD3, and FD6 with $\mathcal{L}_Z^{\text{FCC-ee}}$ reach smaller $\lambda'_{112}/m^2_{\tilde{f}}$ than the CEPC/FCC-ee's ND.
The limit of FD6 reaches $\lambda'_{112}/m^2_{\tilde{f}}=4\times 10^{-15}$ GeV$^{-2}$ at the large mass threshold with $\mathcal{L}_Z^{\text{FCC-ee}}$.

FD1 with $\mathcal{L}_Z^{\text{FCC-ee}}$ has almost the same lower reach as the CEPC/FCC-ee's near detector, and FD8 has the weakest sensitivities among the presented representative far detector designs.
FD3 with $D=50$ m is slightly stronger than the CEPC/FCC-ee's ND at the lower reach for the same integrated luminosity, and shows a potential reach of $\lambda'_{112}/m^2_{\tilde{f}}$ at $1\times 10^{-14}$ GeV$^{-2}$ with $m_{\tilde{\chi}_1^0}\sim 40$ GeV and $\mathcal{L}_Z^{\text{FCC-ee}}$, exceeding that of the CEPC/FCC-ee's ND by a factor of $\sim 2$. 
However, at the upper end of reaches, the CEPC/FCC-ee's ND clearly wins out, which can be explained by the fact that with larger values of $\lambda'_{112}/m^2_{\tilde{f}}$ the light neutralinos decay too fast to reach the far detectors.
On the other hand, even MATHUSLA, the one with the strongest projected reaches among the proposed far detectors at the LHC, is weaker than FD1 by more than one order of magnitude.
This is mainly because of the quite different numbers of $Z-$bosons produced at the LHC and lepton colliders. Besides, the acceptance of MATHUSLA at the LHC is also worse than that of a same sized FD at a lepton collider by one to two orders of magnitude.

Moreover, the lower plot of Fig.~\ref{fig:Z2n1n1:BrLim} shows that if the integrated luminosity can be enhanced to $5\,\mathcal{L}_Z^{\text{FCC-ee}}$, FD6 may reach $\sim 2 \times 10^{-15}$ GeV$^{-2}$ in $\lambda'_{112}/m^2_{\tilde{f}}$.

\section{Conclusions and Outlook}
\label{sec:conslusions}

The LHC will enter the era of HL-LHC in the coming years and is projected to accumulate in total 3 ab$^{-1}$ integrated luminosity by the end of HL-LHC around 2035.
Such a large amount of data would allow for potential discovery of very rare decays of gauge bosons, mesons, etc. 
In light of such possibilities, several proposed far detectors such as MATHUSLA, CODEX-b, FASER, and AL3X at the LHC have been brought up in order to search for LLPs produced from exotic decays of the known SM particles and potential new heavy particles.

As the LHC continues operation, the possibility of building next-generation lepton colliders, working at a series of center-of-mass energies, has been discussed widely in the high-energy physics community. These proposed accelerators are suggested to be $e^- e^+$ colliders including the CEPC and FCC-ee which may work as $Z-$, $W-$ and Higgs factories, etc.

Among all physics goals, it is also interesting to search for the displaced vertices signatures arising from LLPs in the lepton colliders' environment. 
In this study, we have cast a first look of placing a new detector at a position far from the IP at a general $e^- e^+$ collider. 
We develop various designs of such far detectors by varying the locations, volumes, and geometries, and study their sensitivities in three physics scenarios by performing Monte-Carlo simulation: the SM Higgs boson decays to a pair of long-lived scalars $h\rightarrow XX$; the $Z-$boson decays to a HNL and an active neutrino $Z\rightarrow N \nu$; and the $Z-$boson decays to a pair of the lightest neutralinos $Z\rightarrow \tilde{\chi}_1^0 \tilde{\chi}_1^0$ in the context of the RPV-SUSY. 
We compare the limits of such far detectors with those of the default near detectors at the CEPC and FCC-ee and of other proposed LHC far detectors.
Our study has arrived at the following list of conclusions.

\begin{enumerate}
\item At the $e^- e^+$ colliders, the LLPs produced from the Higgs bosons or $Z-$bosons are much less boosted in the forward direction compared to those at the LHC, mainly as a result of the lack of parton distribution inside the electrons and positrons. This comparison is shown in Fig.~\ref{fig:ThetaDistribution}. Consequently, at $e^-e^+$ colliders, a parallel counterpart to FASER which has been approved to be installed at the LHC would be only sensitive to a limited region in the parameter space for the physics scenarios where the LLPs are produced from decays of $Z-$ and Higgs bosons. We therefore do not consider a small detector located at the very forward direction downstream toward the IP at an $e^- e^+$ collider. 

\item We consider 8 different cuboid designs (FD1-FD8) of far detectors each with two benchmark options of its distance ($D$) from the IP. A summary of their setup can be found in Sec.~\ref{sec:fardetectors}.

\item We show further in Fig.~\ref{fig:FidEffRatio} the enhancement of the average decay probabilities as a function of the LLP mass for some representative far detectors in all three physics scenarios, compared to the case with the CEPC's near detector only.

\item We compare the acceptance of displaced vertices for all designs quantitatively in Table~\ref{tab:detectorsefficiencies}, where the average decay probabilities times the proper decay length are presented in the large $c\tau$ limit for LLPs of mass fixed at 1 GeV.

\item In Figs.~\ref{fig:FidEffRatio} - \ref{fig:Z2n1n1:BrLim}, we show the sensitivities for the representative far detector designs: FD1, FD3 , FD5, FD6, and FD8.

\item The different designs allow for understanding the effect of volume, solid angle coverage, distance to the IP, etc., and their interplay, on the sensitivity reaches. In reality, of course, practical considerations must be taken into account. 
However, our results for the preliminary designs  may be a useful reference so that such far detectors could be included into the construction plan of future lepton colliders such as the CEPC and FCC-ee.

\item In general, among all far-detector designs, FD5 and FD6 are expected to have the strongest discovery potential thanks mainly to their gigantic volumes, though the latter might be expensive and thus not realistic. However, a MATHUSLA-sized design such as FD3 and FD4 can already provide a modest and complementary contribution to probing the parameter space of the considered LLP models.

\end{enumerate}

In summary, our study demonstrates that when designing the lepton colliders, the possibility of building an additional far detector might be taken into account in order to achieve further sensitivities on the LLP searches. 

We offer a few further comments below:

\begin{enumerate}
	
\item In this study, we compare the designs according to only three physics scenarios.
It would be interesting to investigate their physics potential and optimize the designs in the context of more theoretical scenarios.
Furthermore, it would be also important to take into account more realistic factors when building such far detectors at a lepton collider, including the availability of the space, the technology and cost of the detectors, the reusing possibility at the SppC/FCC-hh, etc.
We leave them for future studies.

\item In principle, one could perform a sensitivity study for a far detector installed at the ILC which would be operated as a Higgs factory. However, only up to 400 thousand SM Higgs bosons are expected to be produced even with the upgraded luminosity~\cite{Borzumati:2014zxa}, which is below the projected production at the CEPC or FCC-ee (1.14 million cf. Table~\ref{tab:physicscases}).

\end{enumerate}
\medskip

\bigskip
\centerline{\bf Acknowledgements}

\bigskip
We thank Kingman Cheung, Florian Domingo, Oliver Fischer, Martin Hirsch, V\'ictor Mart\'in Lozano, Manqi Ruan, and Abner Soffer for useful discussions.
K.W. acknowledges supports from the Excellent Young Talents Program of the Wuhan University of Technology, the CEPC theory grant (2019-2020) of IHEP, CAS, and the National Natural Science Foundation of China under grant no.~11905162.
Z.S.W. is supported by the Sino-German DFG grant SFB CRC 110 ``Symmetries and the Emergence of Structure in QCD", the Ministry of Science, ICT \& Future Planning of Korea, the Pohang City Government, and the Gyeongsangbuk-do Provincial Government through the Young Scientist Training Asia-Pacific Economic Cooperation program of APCTP. Z.S.W. also thanks the National Center for Theoretical Sciences in Hsinchu, Taiwan for hospitality, where part of this work was conducted.

\bigskip

\bibliography{Refs}
\bibliographystyle{h-physrev5}
\end{document}